\pdfoutput=1
\documentclass[fleqn,usenatbib]{mnras}

\usepackage{newtxtext,newtxmath}

\usepackage[T1]{fontenc}

\usepackage{siunitx}
\usepackage{array}
\usepackage{makecell}
\usepackage{booktabs}
\usepackage[table,xcdraw]{xcolor}
\usepackage{multirow}
\usepackage[normalem]{ulem}
\useunder{\uline}{\ul}{}
\usepackage{pdflscape}
\usepackage{wrapfig}
\usepackage{lscape}
\usepackage{rotating}
\usepackage{etoolbox}
\makeatletter
\patchcmd\@combinedblfloats{\box\@outputbox}{\unvbox\@outputbox}{}{%
   \errmessage{\noexpand\@combinedblfloats could not be patched}%
}%
 \makeatother
\usepackage{graphicx}	
\usepackage{subcaption}
\usepackage{amsmath}	
\usepackage{amssymb}	

\hypersetup{pdfauthor={R. D. Hall},
               pdftitle={On the Feasibility of Intense Radial Velocity Surveys for Earth-twin Discoveries},
               pdfkeywords={methods: data analysis -- methods: statistical -- techniques: radial velocities -- Sun: activity -- stars: activity -- planets and satellites: detection},
               bookmarksnumbered=true}

\newcommand{\programscript}[1]{{\sc #1}}
\newcommand{\MultiNest}{\programscript{MultiNest}}
\newcommand{\PolyChord}{\programscript{PolyChord}}

\title[On the Feasibility of Intense Radial Velocity Surveys for Earth-twin Discoveries]{On the Feasibility of Intense Radial Velocity Surveys for Earth-twin Discoveries}

\author[R. D. Hall et al.]{%
    Richard D. Hall,$^{1}$\thanks{E-mail: rdh47@cam.ac.uk}
    Samantha J. Thompson,$^{1}$
    Will Handley,$^{1,2}$
    Didier Queloz$^{1,3}$
    \\
    $^{1}$Astrophysics Group, Cavendish Laboratory, J. J. Thomson Avenue, Cambridge, CB3 0HE, UK\\
    $^{2}$Kavli Institute for Cosmology, Cambridge, Madingley Road, Cambridge, CB3 0HA, UK\\
    $^{3}$Observatoire Astronomique de l'Universite de Geneve, Chemin des Maillettes 51, CH-1290 Sauverny, Switzerland
}

\date{Accepted XXX. Received YYY; in original form ZZZ}

\pubyear{2018}
\begin{document}
\label{firstpage}
\pagerange{\pageref{firstpage}--\pageref{lastpage}}
\maketitle

\begin{abstract}
This work assesses the potential capability of the next generation of high--precision Radial Velocity (RV) instruments for Earth--twin exoplanet detection. From the perspective of the importance of data sampling, the Terra Hunting Experiment aims to do this through an intense series of nightly RV observations over a long baseline on a carefully selected target list, via the brand--new instrument HARPS3. This paper describes an end-to-end simulation of generating and processing such data to help us better understand the impact of uncharacterised stellar noise in the recovery of Earth-mass planets with orbital periods of the order of many months. We consider full Keplerian systems, realistic simulated stellar noise, instrument white noise, and location--specific weather patterns for our observation schedules. We use Bayesian statistics to assess various planetary models fitted to the synthetic data, and compare the successful planet recovery of the Terra Hunting Experiment schedule with a typical reference survey. We find that the Terra Hunting Experiment can detect Earth--twins in the habitable zones of solar--type stars, in single and multi--planet systems, and in the presence of stellar signals. Also that it out--performs a typical reference survey on accuracy of recovered parameters, and that it performs comparably to an uninterrupted space--based schedule.
\end{abstract}



\section{Introduction}
\label{Introduction}

Ever since the first exoplanet orbiting a main-sequence star was discovered in 1995 \citep{Mayor:1995aa}, the field of exoplanets has witnessed an exponential growth in discoveries, funding, and publications \citep{2015ARA&A..53..409W}. The ``exoplanet zoo'' is constantly growing: planet properties and system architectures are being discovered that redefine our perception of what a normal solar system looks like. However, despite the 3,000 planets found so far, we are yet to confirm the presence of an Earth-twin: a rocky, $1M_{\earth}$ planet orbiting a solar type star in the habitable zone. 

Following the success of the initial radial velocity (RV) surveys, e.g. ELODIE \citep{1996A&AS..119..373B}, HARPS and HARPS--North (HARPS--N) \citep{2003Msngr.114...20M,2012SPIE.8446E..1VC}, wide-field multi--target transit surveys have taken the lead in exoplanet discovery. Programmes such as WASP \citep{2006PASP..118.1407P}, NGTS \citep{2017arXiv171011100W}, Kepler \citep{2010Sci...327..977B} and TESS \citep{2014SPIE.9143E..20R} have, and will, continue to discover thousands of exoplanets. Despite the success of Kepler, it did not provide clear insight into Earth-twin statistics \citep{2016ApJ...830....1K}. RV programmes are still crucial in expanding our knowledge of low--mass, long--period exoplanets.

The recent summary of the state of the field of extreme precision radial velocities by \citet{2016PASP..128f6001F} highlights the milestones required to overcome the \SI{0.1}{\meter\per\second} threshold, the  amplitude of the RV of the Sun due to the Earth. There are a number of upcoming instruments with the goal of finding Earth analogues including ESPRESSO (Echelle SPectrograph for Rocky Exoplanets and Stable Spectroscopic Observations) \citep{2014arXiv1401.5918P}, EXPRES (The EXtreme PREcision Spectrograph) \citep{2016SPIE.9908E..6TJ}, and HARPS3 (High Accuracy Radial Velocity Planet Searcher) \citep{2016SPIE.9908E..6FT}.
Once the instrumental \SI{0.1}{\meter\per\second} accuracy level is met, we will have the capability of finding Earth-twin candidates and be able to provide context for the currently unique configuration of our solar system. 

Breaching this threshold, however, is not only instrument limited. The problem of stellar activity--induced signals is a major barrier in Earth-twin detection. The intrinsic variability of the stellar atmosphere, and our inability to fully correct for the RV signals arising from this, greatly raises the noise floor of our measurements and provides many opportunities for false positive detections. One technique to mitigate this is a sufficiently dense observation schedule over many years. The Terra Hunting Experiment \citep{2016SPIE.9908E..6FT} aims to discover Earth-twins via daily observations for a 10--year programme duration. The high measurement precision of HARPS3 and a radical change in observation schedule proposed by the Terra Hunting Experiment will give observers the best chance yet to detect low mass, long period planets.

This works describes the results of a simulation of various RV observation schedules and uses the Terra Hunting Experiment as a case study for detecting Earth--twins in the face of synthetic stellar activity noise sources.

The essence of this paper is to demonstrate the benefit of optimal sampling of RV measurements and is not a discussion of appropriate stellar correction or mitigation techniques. It is impossible to perfectly model stellar activity, hence any stellar noise model that we include in our analyses will always be imperfect. As an example of this, we aim to show how far an observer can go assuming some imperfect stellar noise correction but with a change in data sampling. We consider a number of solar system architectures and perform a Bayesian analysis of each, with and without the presence of stellar noise. This paper first describes the HARPS3 and Terra Hunting Experiment in Section \ref{sec:HARPS3}, then we discuss how our simulated data is generated including our observation schedules and stellar signals in Section \ref{sec:THE_SIM}. We then discuss the nested sampling software used to analyse our data in Section \ref{sec:POLYCHORD}, present the results of our analysis, initially without stellar signals in the data, Section \ref{sec:gauss_results}. In Section \ref{sec:SOAP} we discuss the origin and impact of stellar signals and we then include the stellar signals in the data and show the results of a constrained (Section \ref{sec:constrained_results}) and unconstrained (Section \ref{sec:unconstrained_results}) analysis. We finish with a discussion of our false--alarm probabilities in Section \ref{sec:FAP_future}.
\section{HARPS3 and The Terra Hunting Experiment}
\label{sec:HARPS3}

HARPS3 is a stabilised, high--resolution spectrograph being built to conduct the Terra Hunting Experiment. Its predecessors, ~HARPS-S and ~HARPS-N, have demonstrated detection capabilities down to \SI{0.92}{\meter\per\second} \citep{2011A&A...534A..58P}. However, since our target detections are of the order \SI{0.10}{\meter\per\second}, it is evident that these programmes still have some way to go if detections of Earth-like planets orbiting Sun-like stars in the habitable zone are to become possible. HARPS3 will be a close--copy of the other two HARPS instruments but with some improvements primarily in the detector unit \citep{harps3_CCD_spie} and calibration procedures. 

The Terra Hunting Experiment is a 10-year observation programme that will target bright, nearby Sun-like stars in an effort to find a sample of Earth-twins for further analysis and detailed characterisation. It will obtain RV data on a minimum of 40 stars, where each star will be observed every night possible for 10 years. These  data sets of unprecedented sampling intensity offer the best chance of finding the small signal of an Earth-mass planet. Given a late G$6$ dwarf target, a planet at 200-400 day period will likely lie in the habitable zone, and if they have a mass of around $1M_{\earth}$ they will have a RV semi-amplitude of around \SI{0.1}{\meter\per\second}. 

From \citet{2016SPIE.9908E..6FT}, we predict that HARPS3 will have a photometric measurement error of $1\sigma = \SI{0.3}{\meter\per\second}$ for a $V=7.5$ G6 star and hence we use this target as our standard star for our simulations. HARPS3 will also exhibit  systematic effects from various sources including temperature variations and detector non--uniformities amongst others. For this work we assume these have been characterised to well below the photometric measurement error.  We also assume that any target is visible for observation for at least 6 months of the year to ensure we obtain appropriate phase coverage of any potential planets in habitable--zone orbits, i.e. we will obtain continuous data over at least one half of an Earth--twin orbit in a single season, and will have full orbital phase coverage of most planets after a $10$--year survey, unless that planet happened to orbit at exactly $365$ days. 


\section{Modelling Radial Velocities}
\label{sec:THE_SIM}

In this section we outline the procedure used to generate an RV series based on the proposed Terra Hunting Experiment observation schedule and two other schedules. 

\subsection{The Simulation Programme}
\label{sec:sim_programme}

The procedure we use to generate an RV series is as follows: 

\begin{itemize}
\item Define a set of solar system architectures.
\item Define a set of observation schedule timestamps.
\item Use the schedules to create an RV time series for each system.
\item Factor in the seasonal and weather variations associated with each schedule.
\item Add noise sources to our observations:
\begin{enumerate}
\item Measurement white noise 
\item RVs from features rotating on the surface of the star.
\end{enumerate}
\item Fit models (different number of planets) to the data.
\item Use Bayesian analysis and nested sampling to assess which model best describes the data.
\item Examine the posteriors of the favoured models and compare them to the injected planets.
\end{itemize}

\subsection{Defining Keplerian Orbits}
\label{sect:kepler}
We model planetary RV curves by adding Keplerian signals over a ten year window with Kepler's law of periods to obtain the RV semi-amplitude, $K_p$, of the star from a given planet p, \citep{web_RV}, 
\begin{align}
    K_p &= \left[\frac{2\pi G}{P_p}\right]^{1/3} \frac{m_p \sin(i)}{m_s^{2/3}} \frac{1}{\sqrt[]{1-e_p^2}}, 
    \label{eqn:kepler2}
\end{align}
where $G$ is the gravitational constant, \(P_p\) is the orbital period of the \(p\)th planet, $m_p$ is the mass of the \(p\)th planet, $m_s$ is the mass of the star, $i$ is the inclination of system relative to the observer and \(e_p\) is the eccentricity of the \(p\)th planet.

We then formulate the total RV of the star, $v(t_n, j)$,  at time \(t_n\) from \(N_p\) planets by creating a sum of Keplerians, one for each planet, 
\begin{equation}
    v(t_n, j) = V_j - \sum_{p=1}^{N_p} K_p [\sin(\frac{2\pi}{P_{p}}t_{n} + \varpi_p) + e_p\sin\varpi_p],
    \label{eqn:RV}
\end{equation}
where \(V_j\) is the systemic velocity relative to the observer, \(K_p\) is the RV semi-amplitude of the star due to the \(p\)th planet from equation \ref{eqn:kepler2}, \(\varpi_p\) is the longitude of periastron of the \(p\)th planet, and $f_{n,p}$ is the true anomaly of the \(p\)th planet.

For our purposes, we simplify the above by setting all eccentricities to zero, i.e. circular orbits, and set the inclination to 90 degrees such that $\sin(i)=1$. Now with circular orbits, we can term the {\em longitude of periastron} to be the orbital phase of the planet, or how much of the orbit that planet has completed at the time of the first observation. We also assume that the barycentric RV correction has already been applied to our data, which sets $V_j$ to zero. We therefore have just three parameters per planet to fit: RV semi-amplitude, K (\SI{}{\meter\per\second}), orbital period, P (days), and phase, $\varpi$ (radians). Lastly, we have assumed that planetary interactions are negligible on the time-scale of $10$ years. Our largest planetary interaction considered in this simulation is between an Earth--twin and a gas giant. The difference of the RV of a Solar--type star with and without the force component of a Jupiter--mass planet we calculate to be no more than \SI{0.00004}{\meter\per\second} over the period of orbit of the gas giant, negligible for our considerations. 

We justify these simplifications to first assess the best-case scenario of our observations and secondly to increase the speed of our analysis. HARPS3 will primarily be looking for planets that are potentially habitable and a large eccentricity is thought to impart great temperature fluctuations throughout the planet's orbit which may or may not be favourable for life \citep{2017arXiv171001405W}. However, an eccentric planet can have a larger maximum RV than a circular planet which would make our detections easier, but would lengthen our analysis and increase the complexity of our model. As this is a preliminary study we opted for the simpler case and therefore we conservatively assume a large eccentricity is undesirable, and a small eccentricity is negligible so we force $e=0$ for all planets. 

Thus all of our RV series comprise a dataset ${D=\{(t_n,v_n\pm\sigma_n),n=1\ldots N\}}$ of velocity measurements $v_n$, their associated errors $\sigma_n$ and times $t_n$. 

To get a better assessment of the performance of each observation schedule, we created four planetary systems with different architectures. In each case the mass of the star was $0.8M_{\sun}$, and the solar systems are detailed in Table \ref{table:RV}. Our first is a lone $1M_{\earth}$ planet orbiting at 293 days, a period chosen to be out of sync with our 6 month repeating schedules to ensure full phase coverage. Secondly we have a near Solar System analogue: A $200M_{\earth}$ gas giant at 2953 days, accompanied by a Venus--analogue at 197 days and an Earth-twin at 293 days. Our third system is three $1M_{\earth}$ planets at 101, 197, 293 days. Lastly we have a null--case test, a star with no planetary companions where the measurements would be of just instrument and stellar noise. We chose to place to the planets on orbits prime numbers of $101$, $197$, $293$, and $2953$ days respectively to avoid multiples of our window function.


\subsection{Observation Schedules, Weather, and Solar Systems}

We compared three observation schedule types, Terra Hunting, Reference, and Space, and applied each one on the chosen solar systems, see Table \ref{table:obs_schedule} for details. 

Our first schedule is that of the Terra Hunting Experiment. This proposed schedule consists of one observation per target per night. In our case our targets are visible for 6 months, hence we have 180 days of possible observations per calendar year.

The second schedule serves as our frame of reference as it has already proven highly successful in planet discovery on a similar instrument to HARPS3. This schedule is a typical GTO schedule \citep{2011A&A...525A.140D} for the HARPS--N instrument and would form the basis of typical low-mass exoplanet survey. It consists of ten observations per month, for six months a year, totalling 60 possible observations a year.

The last schedule is an idealised schedule with one observation per 24 hour period with no interruptions. It ignores seasonal and weather downtime and also represents a schedule one might obtain from space. For a space--based spectrograph operating a programme similar to the Terra Hunting Experiment, this schedule could be easily realised or even exceeded.

We denote these schedules as THE, REF, and Space respectively, and append the duration of the schedule in years. For example, THE\_$10$ refers to a $10$--year Terra Hunting schedule.

\begin{table}
    \caption{A list of planetary systems modelled in the RV simulation. System 4 is the null--case and contains no planets.}
    \centering
    \begin{tabular}{crrr}
        \toprule
        System & Planet Mass / $M_{\earth}$ &  Period / days & RV  /  \SI{}{\meter\per\second} \\
        \midrule
        1		& 1.00 	& 293 		& 0.11\smallskip\\
        2		& 0.82	& 197		& 0.11\\
        & 1.00	& 293		& 0.11\\
        & 200.00	& 2953		& 10.34\smallskip\\
        3		& 1.00 	& 101 		& 0.16\\
        & 1.00	& 197 		& 0.13\\
        & 1.00	& 293 		& 0.11\smallskip\\
        4		& N/A		& N/A 	& N/A\\
        \bottomrule
    \end{tabular}

    \label{table:RV}
\end{table}

\begin{figure}
    \includegraphics[width=\columnwidth]{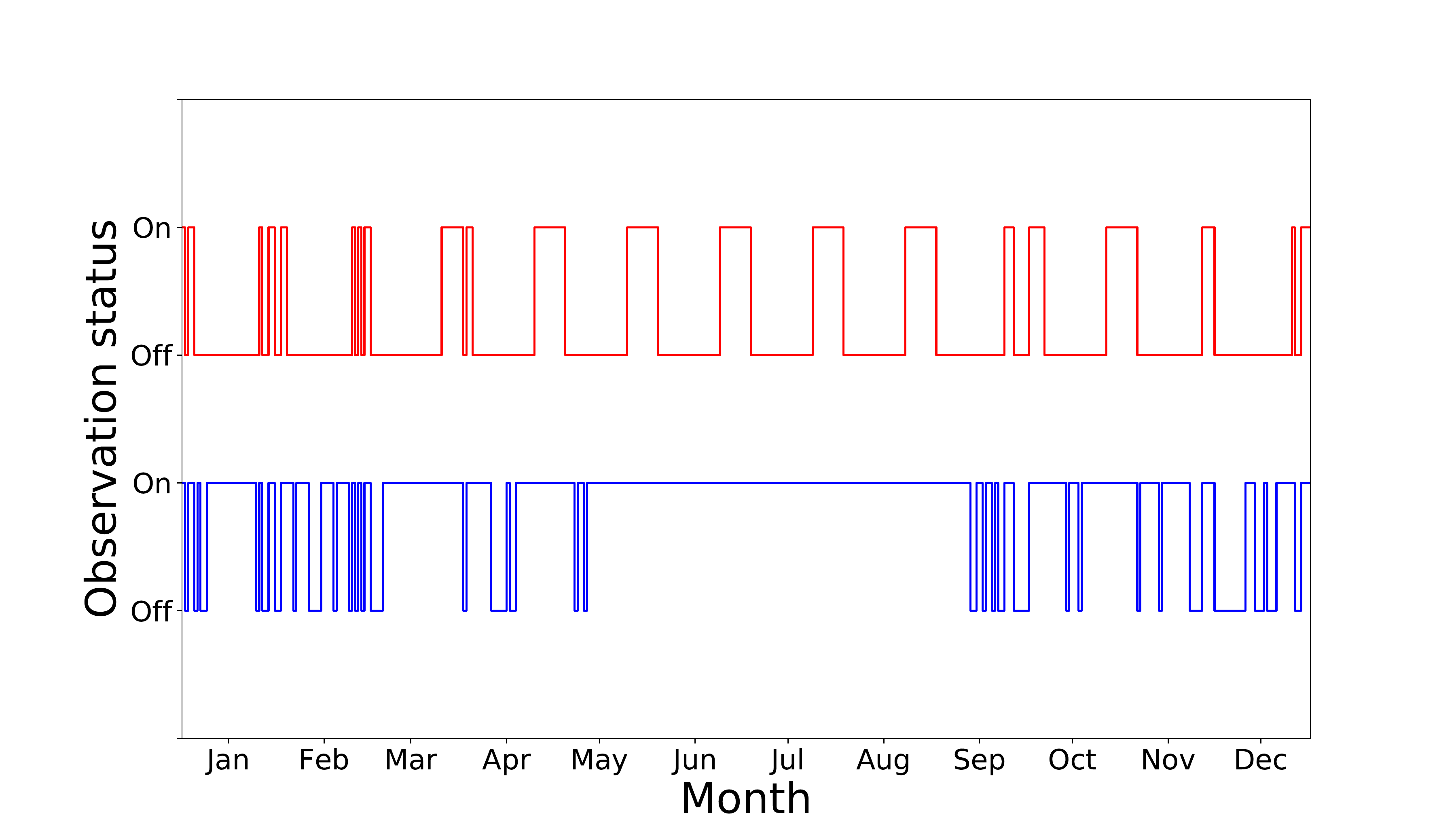}
    \centering
    \caption{A 12-month section of the weather mask as seen by the reference (top) and HARPS3 (bottom) schedules.}
    \label{fig:weather_mask}
\end{figure}

We then account for the seasonal weather variations by extrapolating from a database of observations from La Palma that spans a $5.5$ year period.  The database is structured as `the number of hours per night of weather downtime', not the actual timestamps of the weather itself. Nevertheless, this is enough to see seasonal variation of weather patterns throughout the calendar year, see Fig. ~\ref{fig:weather_mask}. We create a binary mask for each time--stamp and make the assumption that if over half the night is bad, we cancel the observation and give the the binary mask a value of $0$ (`off') for that time--stamp. 

We have assumed that our targets are visible for $6$ months of every year, but we can `choose' when in the calendar year the target is available and can asses the planetary recovery rates for schedules based in the summer or in the winter. For the purposes of this study of assessing the best case scenario, we have only analysed the summer cases. Using the threshold we keep around 94 percent (1695/1800) of observations for a star whose six month observation window is centred over the summer, and around 74 percent (1348/1800) for one in the winter. 

Lastly, we add white noise to represent photon shot noise on each measurement. We draw a random value from a normal distribution of $1\sigma = \SI{0.3}{\meter\per\second}$ and add it to each measurement in each series to represent the photon shot noise for HARPS3. As mentioned in Section \ref{sec:HARPS3}, early optical simulations have anticipated this is reasonable estimate for the noise for a 20--minute exposure of a quiet,  $~V_{mag}=7.5$ star \citep{2016SPIE.9908E..6FT}. 

\begin{table}
    \caption{A list of observation schedules used in the simulation with the total number of observations possible for each schedule per season.}
    \centering
    \begin{tabular}{cccc}
        \toprule
        Survey & Season & Duration / years &  Observations \\
        \midrule
        THE		& Summer 	& 10.0 	& 1695 	\\
        & Summer 	& 5.0 	& 847 	\\
        Reference		& Summer 	& 10.0 	& 507 	\\
        & Summer 	& 5.0 	& 259 	\\
        Space		& All-year	& 10.0 	& 3650 	\\
        & All-year	& 5.0 	& 1825 	\\
        \bottomrule
    \end{tabular}

    \label{table:obs_schedule}
\end{table}


\section{Bayesian Inference and Nested Sampling}
\label{sec:POLYCHORD}

\subsection{Bayes Theorem}

This work makes use of Bayesian analysis to fit models to data. 
In the context of RV data, Bayesian model selection allows for the rigorous calculation of posterior probabilities of models with different numbers of planets. Comparing how well these models fit the data can be used when considering different observation strategies. It is expected that as you increase the quantity and quality of data, by varying the observation schedule, the true model will become favoured over the false models. In our case `model' refers to the number of planets we are trying to fit to the data. For a given series of measurements (one of our observation schedules applied to one of the four solar systems) we fit a $0$, $1$, $2$, $3$, and $4$ Keplerian planet model.

We follow the notation and methodology of \cite{2011MNRAS.415.3462F}. Given data $D$ and a model $H$ with parameters $\Theta$, Bayes' theorem states:
\begin{equation}
    \mathrm{P}(\Theta | D, H) = \frac{\mathrm{P}(D|\Theta, H) \mathrm{P}(\Theta,H)}{\mathrm{P}(D|H)} \:\leftrightarrow\: \mathcal{P}(\Theta) = \frac{\mathcal{L}(\Theta)\pi(\Theta)}{\mathcal{Z}}
    \label{eqn:bayes}
\end{equation}
where 
$\mathcal{P}(\Theta)$ is termed the posterior distribution of the parameters,
$\mathcal{L}(\Theta)$ the likelihood,
$\pi(\Theta)$ the prior and
$\mathcal{Z}$ the Bayesian evidence. 
After specifying a model via its likelihood and prior, one may numerically sample the posterior and compute the evidence using nested sampling \citep{skilling2006}.

The posterior distribution is used to perform {\em parameter estimation}, namely the quantification by $\mathcal{P}(\Theta)$ of our knowledge of a model's parameters in light of the data and our prior assumptions.

The evidence $\mathcal{Z}$ is used to perform {\em model comparison}. Applying Bayes' theorem again to a sequence of models $\{H_0,H_1,\ldots\}$:
\begin{equation}
    \mathrm{P}(H_i|D) = \frac{\mathrm{P}(D|H_i)P(H_i)}{P(D)} = \frac{\mathcal{Z}_i\phi_i}{\sum_k\mathcal{Z}_k\phi_k}
\end{equation}
we can see that the evidences $\mathcal{Z}_i=\mathrm{P}(D|H_i)$ along with the model priors $\phi_i=\mathrm{P}(H_i)$ can be used to infer the relative likelihood of a model $H_i$ within the set of models. The model priors are typically taken to be uniform, and one can compare the relative probabilities of two models via a Bayes factor:
\begin{equation}
    \label{eqn:bayes2}
    R = \frac{\mathrm{P}(H_1|D)}{\mathrm{P}(H_0|D)} = \frac{\mathrm{P}(D|H_1) \mathrm{P}(H_1)}{\mathrm{P}(D|H_0) \mathrm{P}(H_0)} = \frac{\mathcal{Z}_1\phi_1}{\mathcal{Z}_0\phi_0}
\end{equation}
The natural logarithm of $R$ provides us with a convenient measure of what constitutes a significant difference between two models and is summarised in Table~\ref{table:bayes_evidence}.

To perform nested sampling, we use the implementation provided by \PolyChord{} \citep{2015MNRAS.450L..61H,2015MNRAS.453.4384H}. In the past, the nested sampling algorithm \MultiNest{} \citep{2008MNRAS.384..449F,2009MNRAS.398.1601F,2013arXiv1306.2144F} has been successfully applied to both real and synthetic exoplanet RV data \citep{2011MNRAS.415.3462F},  and was also used in the recent RV Challenge \citep{2016A&A...593A...5D, 2017A&A...598A.133D}. {}\PolyChord{} is a successor to \MultiNest{}, designed to work more efficiently with a higher number of dimensions. \PolyChord{} was chosen over \MultiNest{} for reasons of scalability and future-proofing with reliable updates and fixes in the foreseeable future. Whilst \MultiNest{} is faster than \PolyChord{} for sampling up to approximately $64$ dimensional spherical Gaussian posteriors, this crossover threshold is significantly lowered for more complicated posterior shapes. For our exoplanet distributions, the crossover was found to be between $10$ and $20$ dimensions. Preliminary runtime tests of \MultiNest{} were between $2$ and $3$ times longer on average.  In these tests we also examined the posteriors of simple circular orbit parameters of some test--case planets. We found that as the number of planets were increased, \MultiNest{} results started to deviate from the true values, but were recovered correctly by \PolyChord{}.  Moreover, \MultiNest{} requires a relatively high minimal number of live points $n_\mathrm{live}\sim400$, whilst \PolyChord{} produces accurate posteriors with far lower numbers $n_\mathrm{live}\sim5n_\mathrm{dims}$. Running the analysis of all observation schedules applied to one solar system using \PolyChord{} took at most $57$ hours on a single core of a $2.90$GHz Intel Xeon E$5$--$2690$. There is capacity to speed this up by a factor of $10-100$ for future pipelines by re-writing the likelihoods in C/C++ rather than Python.

\begin{table}
    \caption{The Jeffreys scale for interpretation of model probabilities \citep{jeffreys1983theory}.}
    \centering
    \begin{tabular}{cccc}
        \toprule
        |$\Delta \ln R$| & Odds & Probability & Remarks\\
        \midrule
        < 1.0 & $\lesssim$ 3:1 &  0.750 & Inconclusive \\
        1.0	& $\sim$3:1 & 0.750 & Weak Evidence \\
        2.5	& $\sim$12:1 & 0.923 & Moderate Evidence \\
        5	& $\sim$150:1 & 0.993 & Strong Evidence \\
        \bottomrule
    \end{tabular}

    \label{table:bayes_evidence}
\end{table}


\subsection{The Likelihood}
\label{sect:likelihood}

For our first set of results, we omit the stellar noise from the RV data and only include the purely random Gaussian component of the noise. As such our model treatment of the noise as being purely Gaussian and independent is correct, so we can describe the likelihood as minimising a least--squares fit. In our later analysis we add in the structured, correlated, and quasi--periodic stellar signals but we do not change our likelihood function, i.e. in this case we are incorrectly modelling the noise structure of our data. We know we will never have a full description of the stellar noise, however we want to demonstrate how far optimal sampling can get you in the common situation of incomplete noise modelling.

In the cases where we have stellar signals but have not modelled them, we are then assessing the impact that these uncharacterised signals have on our ability to recover the Earth--twin candidates and how that ability changes with different schedules and sampling. It is predicted that for smaller planetary signals, the stellar noise will dominate and induce false positives in the posterior distributions. If $v(t;\Theta)$, our model, is the predicted RV function for a set of orbital parameters $\Theta$, then given the aforementioned assumptions the likelihood is simply \citep{mackay2003information}:

\begin{equation}
    \label{eqn:likelihood}
    \ln\mathcal{L}(\Theta) = \sum_i -\ln\sqrt{2\pi\sigma_i^2} - \frac{1}{2\sigma_i^2 }[v(t_i;\Theta) -v_i]^2.
\end{equation}
As mentioned in Section~\ref{sect:kepler}, we simplify our orbital parameters $\Theta$ by forcing the system inclination, barycentric velocity, and eccentricities to zero. This gives us just three parameters per planet $p$ to fit: orbital period $P_p$, RV semi-amplitude $K_p$ and orbital phase $\varpi_p$.
Using equation~\eqref{eqn:RV}, $v(t;\Theta)$ is a sum of sinusoids, with one for each of $N_p$ planets in the system:
\begin{equation}
    v(t;\Theta) = \sum_{p=1}^{N_p} K_p \sin\left[ \frac{2\pi t}{P_p} + \varpi_p\right].
    \label{eqn:polychord_model}
\end{equation}


\subsection{The Priors}
We define our priors in Table~\ref{table:priors}. We choose a lower bound on each period $P$ of $10$ days to eliminate any solutions that are affected by short period stellar noise and the upper bound is set to be just longer than the length of our survey duration to allow the solution to find up to a complete period of a Jupiter--analogue. For our study of looking for long period planets, an alternative and possibly more reliable lower bound for the period could have been set to ensure we avoid the rotation period of the star, e.g. $50$ days. However, to keep this investigation more general we use a lower prior on the period. This also allows for the investigation of possible sources of false positives arising from stellar signals. For the RV semi-amplitude $K$, the range is set for practicality to be a hot giant planet at $2000$\SI{}{\meter\per\second}. The orbital phase is simply uniform in $[0:2\pi]$. In order to break the $N_p!$ switching degeneracy between sets of planet parameters \cite{2007MNRAS.374.1321G}, we choose to impose a prior constraint on the periods such that first planet has the smallest period, followed by the second and so on. This ordered prior constraint could still be vulnerable to signals of very similar periods and amplitudes, e.g. two signals of $10.00$ days at $1$\SI{}{\meter\per\second} and $9.99$ days at $1$\SI{}{\meter\per\second} could be fitted by a single curve of $10$ days at $2$\SI{}{\meter\per\second}, but it reduces the degeneracy from $N!$ to just $N$.

\begin{table}
    \caption{Specification of prior probability distributions on the parameters.} 
    \centering
    \begin{tabular}{cccc}
        \toprule
        Parameter & Prior  & Lower Bound & Upper Bound\\
        \midrule
        $P$ (days) & Log Sorted Uniform & 10 & 4000 \\
        $K$ (\SI{}{\meter\per\second}) & Log Uniform & 0.05 & 2000 \\
        $\varpi$ (radians) & Uniform & 0 & 2$\pi$ \\
        \bottomrule
    \end{tabular}
    \label{table:priors}
\end{table}

\subsection{Using PolyChord}
For each of our series of measurements, we loop through the models of $N_p = 0, 1, 2, 3$, and $4$ in turn, and record the evidence of that model. We limit our search to 4 planets to save on computational expense. The high-dimensional posterior probability distribution is well visualised in a violin plot, examples of which are given in Section~\ref{sec:constrained_results}.


\section{Results from data with only Gaussian noise and planets}
\label{sec:gauss_results}

Our first set of results are from fitting planetary models using RV measurements that only have Gaussian noise, i.e. our data sets only contain planetary RVs (see Table \ref{table:RV} for system descriptions) and Gaussian noise at the timestamps of the schedules. Each of these measurement series were fed into \PolyChord{} and models of $0$--$4$ planets were fitted. 

\subsection{System 1 Results}
\label{subsec:sys1_nosoap_results}

Table \ref{tab:NO_SOAP_sys1} shows the results of system 1 for each schedule. Here, each schedule favours the true 1--planet model with good significance over the nearest models. Only the 5--year reference schedule has a moderate significance compared to a false model. The parameters of these 1--planet models are shown in the right--hand side of Table \ref{tab:NO_SOAP_sys1} to see if any of the results favoured the true $293$--day, \SI{0.11}{\meter\per\second} signal. Here it clear that for a Gaussian--noise only data set, and for likelihood function that correctly assumes only Gaussian noise on the measurements, all schedule results have returned reasonable parameter estimates. All schedules show an RV amplitude which sits within $1\sigma$ of error from the true value. The estimates and errors on the periods scale well with the number of data points in each time series. From REF\_$10$ to THE\_$10$ there is only a three--fold increase in the number of data points, however the accuracy of the parameter estimates are similar whilst the error has decreased by a factor of $60$. This simple case highlights the difficulty of finding low--amplitude RV signals buried in white noise, and the necessity of good sampling and intense measurements to effectively bin down the noise. 

\subsection{System 2 Results}
\label{subsec:sys2_nosoap_results}

Similarly as for section \ref{subsec:sys1_nosoap_results}, we present the results of system 2, the close--copy of our own solar system, in Table \ref{tab:NO_SOAP_sys2}. Here we have a Venus, Earth, and Jupiter analogue system designed to be a test of finding low amplitude planets hidden in a long--period, high--amplitude signal.

We can see that both the reference schedules ($5$ and $10$ years) have favoured false models. Their detected lower--mass companions are also all false positives indicating this particular cadence may struggle to differentiate a combination of signals given that they were successful for system 1. 

Both of the Terra Hunting schedules favoured the true model, but only THE\_$10$ returned accurate and precise parameters. The THE\_$5$ failed to find the Venus--analogue and instead fitted a 2--year planet with an almost $90\%$ error. 

The two space schedules performed very well with Space\_$5$ having slightly larger errors than THE\_$10$ but similar parameters. Space\_$10$ recovered all planets, mostly within $1\sigma$ of the correct value apart from the Jupiter period but the error estimate is low. These results add strong credence to the growing body of evidence that less dense and uneven sampling strategies are insufficient to accurately and reliably retrieve complex and low amplitude signals in noisy data, for example \citet{2017MNRAS.471L.125R}. 

\subsection{System 3 Results}
\label{subsec:sys3_nosoap_results}
System $3$ is a collection of $3$ Earth--mass planets on three different orbits. This configuration acts as a test of finding and differentiating similar signals superimposed with each other. 

Both of the reference schedule results favour a $4$ planet model, 1 additional planet from the true $3$ planet model. Interestingly, REF\_$5$ recovered the $101$--day period planet with 3 other false--positives, whilst, REF\_$10$ missed that planet but found the $197$--day planet accompanied with 3 different false--positives. For the reference schedule results the errors on the parameters were very large on the false positives, up to $50\%$, whilst the errors on the successfully found planets were much lower. 

Both of the Terra Hunting schedules found all three planets with no additional companions, and have much smaller errors on the parameter estimates. The THE\_$10$ parameter estimates were marginally better and had lower errors on all estimates. 

As with System 2, the Space\_$5$ results again sit between THE\_$5$ and THE\_$10$ in terms of estimates and $1\sigma$ errors, whilst the Space\_$10$ results have the highest confidence and are the most accurate. These results suggest that there could be a plateau in the optimum sampling of data with respect to quality of parameter estimation. The number of data points for REF\_$10$ versus THE\_$5$ is roughly two--thirds, whilst the performance of the latter schedule is drastically improved. Doubling the size of the data again for THE\_$10$ only marginally improves the estimates. The full results are in Table \ref{tab:NO_SOAP_sys3}.

\subsection{System 4 Results}
\label{subsec:sys4_nosoap_results}
System 4 is a null--planet test to see what \PolyChord{} would fit to an essentially empty dataset containing only white--noise. Here the true zero--planet model is strongly favoured both of the reference schedules. It is strongly disfavoured for the Terra Hunting and Space schedules. For these, the likely explanation is that the density of the data has forced some edge--case false--positives to be favoured as they all have amplitudes at the lower bound of our priors, and have periods comparable to aliases/harmonics of our window functions. Also the large errors on these values mean the solution is essentially unconstrained. The full results are in Table \ref{tab:NO_SOAP_sys4}.


\section{Including Stellar Signals}
\label{sec:SOAP}

We now add a second type of noise to our data, RVs from the star itself. This is introduced by time--varying active regions of the stellar surface. Stellar signals manifest in the data as a source of quasi-periodic noise on varying temporal baselines that can be falsely interpreted as planetary signals (false--positives). The stellar RV signals are related to four physical phenomenon, each with a characteristic RV amplitude and period:

\begin{enumerate}
    \item Pressure waves propagating in the convective zone of the star; RVs of \SIrange{0.1}{4}{\meter\per\second} over a few minutes \citep{2011A&A...525A.140D}. 
    \item Granulation from convection at the stellar surface where cool material falls into the star whilst hot material rises from beneath in small cells, RVs of \SIrange{0.1}{4.0}{\meter\per\second} over minutes to days \citep{2011A&A...525A.140D}. 
    \item Short-term stellar activity from rotating surface structures, e.g. sunspots creating an RV asymmetry as they move from the red to the blue-shifted half of the star; RVs of \SI{0.5}{\meter\per\second} over 20 days \citep{2011A&A...528L...9L}. 
    \item Long-term magnetic cycles that drive spot formation rates, rising granules of plasma, and overall brightness; RVs of a few \SI{}{\meter\per\second} over many years \citep{2013A&A...551A.101M}. 
\end{enumerate}

Often, the surface activity features linked to stellar rotation arise as false positives. Any low--mass planet claim would be suspicious if measures to mitigate these signals were not adequate. This is a difficulty made infamous with the claim of a short--period planet orbiting the nearby Alpha Cen B \citep{2012Natur.491..207D}, which was then disputed in \citet{2016MNRAS.456L...6R} where these authors used Gaussian processes to demonstrate the window function and stellar signals likely gave rise to spurious signals mistaken for a planet. 

To mitigate the effects of pressure waves, long exposures ($15-20$ minutes) as often as possible (at least once per night) can be binned down and effectively reduced to white noise lower than our assumed photon shot noise. For example, in \citet{2011A&A...525A.140D} the authors used synthetic data to determine an ideal strategy by comparing different sets of exposures in a given night ($1\times30$min or $3\times10$min for example). They showed that multiple exposures per night, repeated daily for binning over $5-10$ days where possible, reduced the RMS of their RV measurements by up to $30$ percent.

The long period magnetic activity of the stars is more difficult to model. These cycles typically last years and drive the formation rate of surface features. We are most familiar with this phenomenon as our Sun has an 11-year activity cycle. The Terra Hunting Experiment will prioritise targets approaching the minimum of the activity cycle and reject those that show signs of high activity.

There are various other techniques for mitigating the effects of stellar signals which are currently utilised to help disentangle the intrinsic stellar signals from potential planets. One is to look at the cross correlation function (CCF) of the spectrum and a suitable template mask that matches the spectral type of the star. High levels of stellar activity linked to rotating surface features create asymmetries in the CCF and can be measured with the bisector inverse slope (BIS). In \cite{2001A&A...379..279Q}, the authors used this technique to demonstrate that a 3.8 day period, \SI{83}{\meter\per\second} signal on HD 166435 was in fact activity related and not a large planetary companion. However, for low amplitude magnetic cycles and small numbers of sunspots, this process is not accurate enough and cannot fully account for the different sources of activity \citep{2016MNRAS.457.3637H}. 

Other methods of mitigating the impact of stellar variability include a Bayesian model of unknown but present correlated (red) noise within RV datasets \citep{2014MNRAS.437.3540F}. Here the authors suggest that a proposed third planet orbiting GJ$667$C has a large evidence of originating from correlated noise and that additional measurements would be required. Another Bayesian--based red noise model utilised an additional short period exponential decay time--scale term to their RV modelling and found that they could potentially model shorter period and lower amplitude planets as this technique accounted for some of the small--scale stellar jitter \citep{2013A&A...551A..79T}. 

Lastly, the current state--of--the--art technique involves a joint modelling of the planetary and stellar signals via a Gaussian process \citep{2015MNRAS.452.2269R}.  This framework simultaneously models time--series with activity indicators from the bisector velocity spans, line widths, and chromospheric activity indices, to allow the planetary signals to be distinguished. This technique was used in the aforementioned Alpha Cen B planet dispute. We do not use these or any other technique to model the stellar noise but we mention them to justify some of our assumptions in section \ref{subsec:generating_stellar_noise}.

\subsection{Generating Stellar Signals}
\label{subsec:generating_stellar_noise}
\begin{figure}
    \includegraphics[width=\columnwidth]{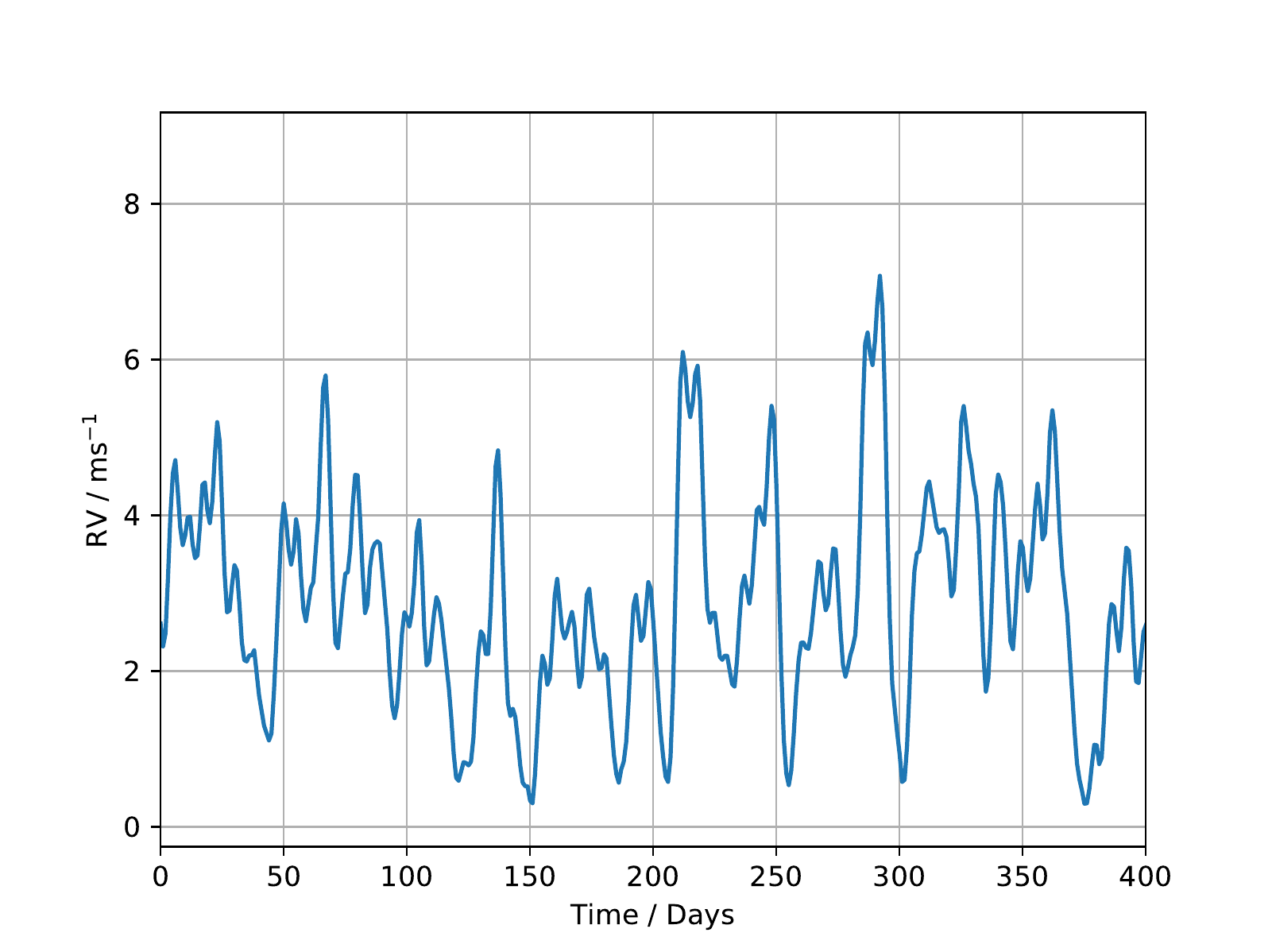}
    \centering
    \caption{An example of the raw RV series from surface features on a quiet solar-type star generated from SOAP$2.0$. As SOAP returns discreet points every $24$ hours we plot a smooth interpolation to better show the structure of the data.}
    \label{fig:SOAP2.0}
\end{figure}

We generate stellar radial velocities from SOAP 2.0 (Stellar Oscillation and Planet) \citep{2014ApJ...796..132D, 2015ascl.soft04021D}, and add these to our planetary signal; see an example in Figure~\ref{fig:SOAP2.0}. SOAP 2.0 estimates the RV variations induced by active regions in the photosphere. With SOAP 2.0 we are able to choose the stellar rotation rate, the effective surface temperature, the stellar mass, the average number of spots present, the coordinates of the active regions, the average lifetime of starspots, and the spot to plage filling factor ratio. 

To tune the various parameters of SOAP2.0, we use the indicator $\log(R'_{HK})$ \citep{1984ApJ...279..763N} to help us define different levels of activity used in this simulation. In \citet{2014ApJ...796..132D, 2015ascl.soft04021D}, the authors describe the parameter values that correspond to different activity levels of a Sun-like star; minimal, low, medium, and high activity. We have chosen parameters to represent a star between minimal and low activity levels with an average spot count of 20, and a spot to plage ratio of 5.

We assume that some correction for stellar noise is possible using the aforementioned techniques, but that no correction process or red noise model is perfect and would leave unaccounted for structured residuals in the data. As this work is not a study of stellar signal mitigation, we have made an assumption that we can approximate the \textit{outcome} of these processes by simply dividing the amplitude of the SOAP RVs by a fixed factor. We justify this approximation by reason that it will leave structure in our data that is characteristic of stellar signals both in amplitude and in time, but with a reduced impact compared to the full stellar noise. To represent optimistic levels of correction that could be possible with the most modern of techniques and a $10$--year RV survey, we use a reduction of the stellar signals by a factor of four. It should be noted that the raw SOAP RV RMS is approximately \SI{4}{\meter\per\second} and is therefore reduced to a RMS of \SI{1}{\meter\per\second} and as such is still many times larger than the RV of the Earth--twin and higher than our instrument photon noise.

We do not claim that this is an accurate representation of a dataset. However we think this is a reasonable representation for our purposes in demonstrating the benefit of different observing schedules in the limit of incomplete noise modelling. We keep our likelihood of a least--squares minimisation of a Keplerian model and now test \PolyChord{} and the schedules in their ability recover the planetary signals in the presence of unmodelled and unknown quasi--periodic stellar RVs.

We acknowledge a few limitations to SOAP$2.0$ which will be addressed in a follow up work. Firstly our stellar noise is generated from approximations of surface features only and does not include pressure waves, granulation or magnetic cycles. The software is not best suited for our application of quiet stars as a spot is always associated with faculae, something we know to be false from resolved observations of our Sun, \citep{2014A&A...569A..38S}. Also, due to computational constraints, we were only able to generate one data point per 24 hours of our schedule. This means our observations have to fall on a regular 24 hour cadence to align with the SOAP RVs. This regular observation pattern is not usually plausible in reality due to constraints from the weather, scheduled and unscheduled instrument maintenance, and demand from other users. We partly addressed this by limiting our prior on period to 10--days and up, such that the 24--hour cadence and its major harmonics will not be a possible solution for our parameter estimation.

Also, as SOAP2.0 cannot simulate long--term magnetic cycles our average spot number and active region size maintains constant throughout our simulation. This would not present an issue if we were simulating small data series representing $100$ days or so, but for a $10$ year simulation of a Sun--like star we would expect around one complete magnetic solar cycle. During minimum activity of the Sun, we see $4-5$ years of very low, near flat activity levels before it ramps up. A programme like the Terra Hunting Experiment could favour targets exhibiting low activity levels in the hope of obtaining a 5--year long series of quiet measurements, and then move on to a new target once the activity rose above some threshold. Hence we have also included $5$--year subsets of our schedules to represent this possible strategy, where the noise is roughly constant and our assumption that we can ignore magnetic cycles is valid. For future work, we aim to either modify the original SOAP$2.0$ code to include an activity level variance as a function of time, or generate sub-sets of RVs with different activity parameter values, and interpolate and stitch them with a Gaussian Process.

\begin{figure}
    \includegraphics[width=\columnwidth]{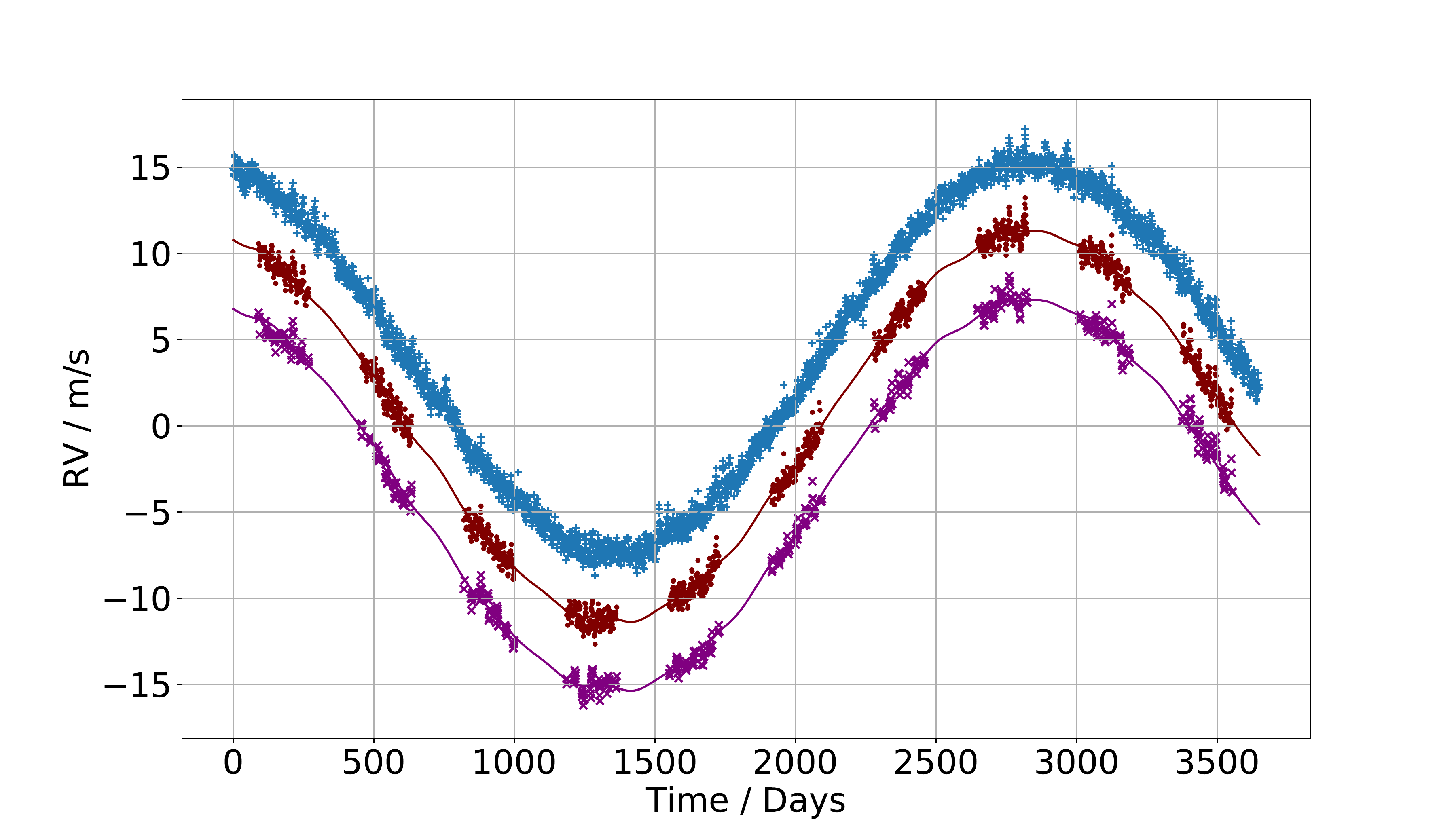}
    \centering
    \caption{A $10$--year simulation of the RV series of System $2$ (see Table~\ref{table:RV}) including SOAP$2.0$ and Gaussian noise, observed with a typical Terra Hunting summer schedule (middle), Reference summer schedule (bottom) and the uninterrupted (space) data (top). The data sets are staggered vertically for clarity, and the solid lines through each are the pure RV curves from the planetary models.}
    \label{fig:rv_10yr}
\end{figure}

\begin{figure}
    \includegraphics[width=\columnwidth]{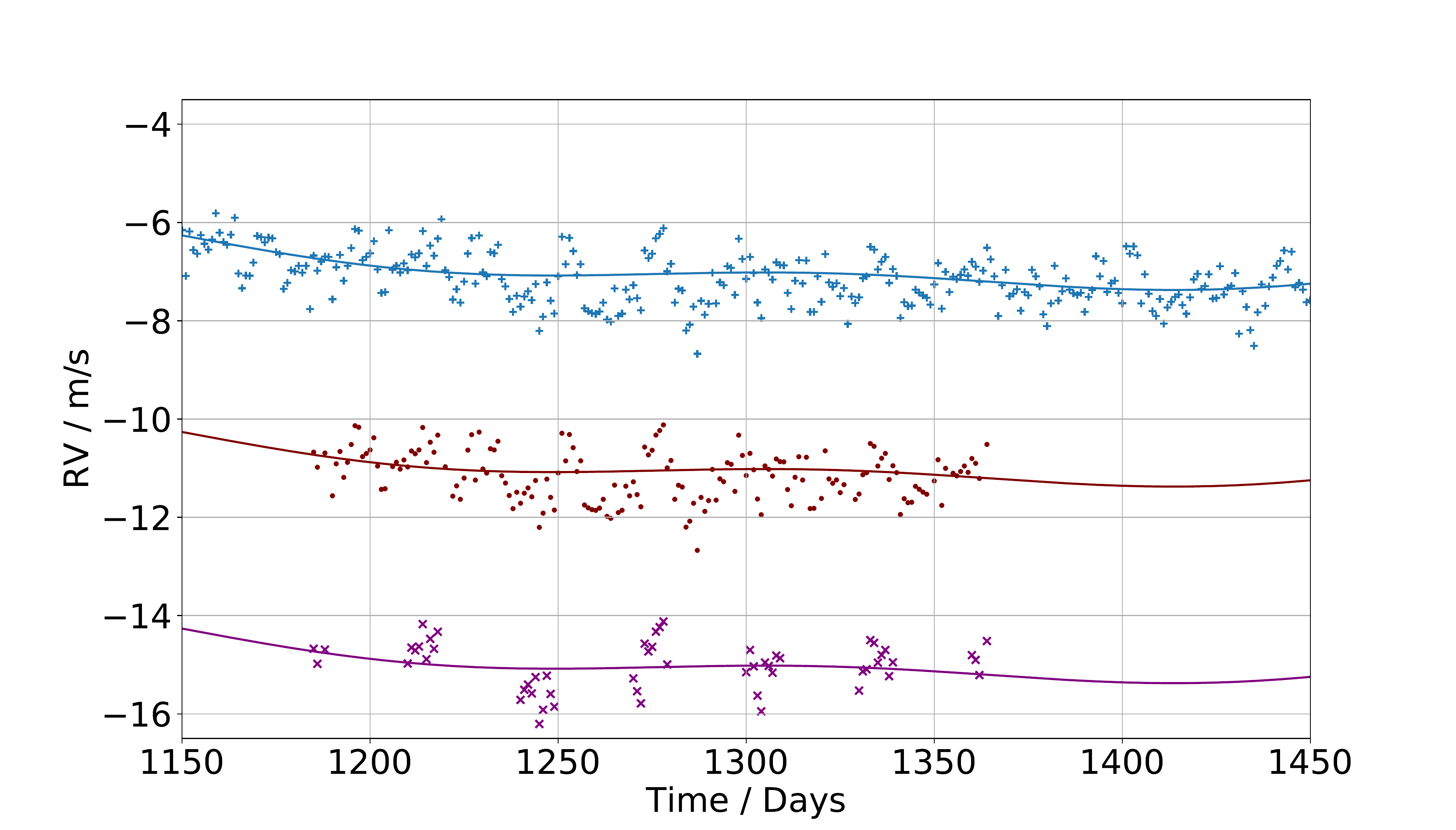}
    \centering
    \caption{A zoom of Fig~\ref{fig:rv_10yr} showing the different observation schedules and demonstrating their relative density.}
    \label{fig:rv_10yr_zoom}
\end{figure}

An example of a System 2 RV series is shown in Fig. ~\ref{fig:rv_10yr} and a zoomed section in Fig. ~\ref{fig:rv_10yr_zoom}. In these figures, the solid line running through the data sets is the noise-free planetary RV model. A large single period of the Jupiter analogue is obvious, but it is the smaller 291 and 197 day period oscillations from the rocky planets that HARPS3 is searching for. Also worth noting is a $25$ day signal visible in the space and Terra Hunting data which originates from a large sunspot group. 



\section{Constrained model Results from data with Stellar Noise, Gaussian noise, and planets}
\label{sec:constrained_results}

As detailed in section \ref{sec:gauss_results}, we now itemise the results per Solar System where the data now includes signals from stellar rotating surface features, and comment on the results. The full tabulated results are in the appended Table  \ref{table:soap_constrainted_full_results}. In this section we have constrained \PolyChord{} to search only for the number planets described by the true model for each solar system, i.e. for system 1 we only allow $N_p = 1$. We later describe the results in an unconstrained model search.

\subsection{System 1 Results}
\label{subsec:sys1_soap_forced}

We force \PolyChord{} to only search for one planet within these datasets to see if it can successfully recover the lone Earth--twin with stellar noise present. 

The REF\_$5$ schedule failed to find the Earth--twin and instead fitted a false positive at $210.13 \pm 69.27$ days with an amplitude of \SI{0.16}{\meter\per\second}. This period could be an alias of the combination of a $30$--day and $180$--day baseline within the schedule. REF\_$10$, however, performed slightly better as the parameters are all within $1\sigma$ of the correct value, but the error is very large. THE\_$5$ found a planet very close to the true parameters with small errors, but the phase was incorrect by almost $\pi/6$, an effect we attribute to the inclusion of structured noise within the measurements. THE\_$10$ performed well and found the correct period and amplitude with small errors. As per the $5$ year schedule the phase was incorrect but this time only by $\pi/12$. The space--based schedules again improved on the THE schedules, with good estimates for the $10$ year survey. However in both cases the phase was still underestimated by a non--negligible amount. 

These results indicate that even with inadequate stellar modelling it is still possible to detect a \SI{0.10}{\meter\per\second} RV signal within larger structured noise. However the presence of structured and quasi--periodic signals appears to affect the placement of that signal at the correct phase. 

\subsection{System 2 Results}
\label{subsec:sys2_soap_forced}

We now force \PolyChord{} to a $3$--planet solution and assess whether the presence of stellar signals prevent an Earth--twin detection that has low and high amplitude companions. 

All schedules recovered the $2953$ day, \SI{10.34}{\meter\per\second} signal from the Jupiter analogue, even the $5$ year surveys which only cover just over half of the orbit returned good estimates. However both of the reference schedules failed to identify the other two companions with the REF\_$5$ schedule returning a planet with a period of $267.74\pm295.75$ days, a $111\%$ error. REF\_$10$ also did not recover any true companions but did find a $27.50\pm16.94$ day signal which is very close to the $25$--day stellar rotation rate introduced by SOAP$2.0$.

THE\_$5$ found the Venus analogue but, as with system 1, placed it out of phase by $\pi/6$. The third planet found by THE\_$5$ is also very similar to the stellar rotation rate at $20.88\pm2.05$ days. THE\_$10$ successfully found the Earth twin at the correct period and phase, but underestimated the RV semi amplitude at \SI[separate-uncertainty = true]{0.08\pm0.01}{\meter\per\second}. The third planet from THE\_$10$ is an $80$ day period false--positive at the lower limit of the RV prior, a result we attribute to \PolyChord{} over fitting the noise due to an inadequate likelihood function.

Lastly, Space\_$5$ found the Earth--twin successfully with parameter estimates slightly broader than THE\_$10$, it also found the Venus analogue with good estimates for the period and RV, but again placed it slightly out of phase. Space\_$10$ found all three planets with good parameter estimates and low errors.

We can attribute the failures of the reference schedules to inadequate sampling of the finer structure of the SOAP RV data, which under this poor resolution allows degenerate solutions to be selected by \PolyChord{} and provides these false--positives. However with the denser sampling from the Terra Hunting and Space surveys, we have enough data to reduce the degeneracy to resolve the stellar signals and allows for the underlying and consistently repeating planetary signals to be recovered.

\subsection{System 3 Results}
\label{subsec:sys3_soap_forced}

Again we constrict \PolyChord{} to a 3 planet solution as a test of its ability to find $3$ similar, low--amplitude signals within the structured stellar noise.

Both reference schedules failed to detect any of the true planets and most of the parameters have very large errors associated with each parameter. Some of the false--positives could be attributed to harmonics of the stellar rotation rate, such as a $75.25\pm11.29$ day planet from REF\_$5$ or a $51.32\pm24.55$ day planet from REF\_$10$. REF\_$10$ also found a $342.78\pm133.52$ day signal which is very close to the $365$ day cadence underlying all of the ground--based surveys. 

The Terra Hunting surveys performed well. Both schedules found all three Earth--mass planets at the correct periods and phases within $2\sigma$, and mostly the RV amplitudes with $3\sigma$. In both the $5$ and $10$ year cases, the Earth--twin amplitude was overestimated at \SI[separate-uncertainty = true]{0.14\pm0.01}{\meter\per\second}. 

The Space\_$5$ outperformed THE\_$10$ for the first time as the errors on all parameters were an order of magnitude lower, with the parameter estimates also being correct. Interestingly Space\_$10$ failed to find the $197$ day planet, and instead found the stellar rotation rate as the lowest period planet. This is likely due to the fact that $10$ years of unbroken measurements are sampling the stellar activity signals modulated by its rotation period very well, i.e. the evidence is higher for this signal, whereas having some gaps from the ground--based surveys can still adequately sample the complete orbital phase of the planets, but misses out half the revolutions of the star compared to the ground--based case. The full results are in Table \ref{table:soap_constrainted_full_results}.

\subsection{System 4 Results}
\label{subsec:sys4_soap_forced}

We can not ask \PolyChord{} to fit zero planets to the data, only compare the likelihood of the data not being described by the model versus the likelihoods of other. Complete System 4 results are shown in the next section where we do not constrain the number of planets for each system, and thus model comparison is possible.

\subsection{Posterior Visualisation and Phase folding}
\begin{figure*}
    \includegraphics[width=\textwidth]{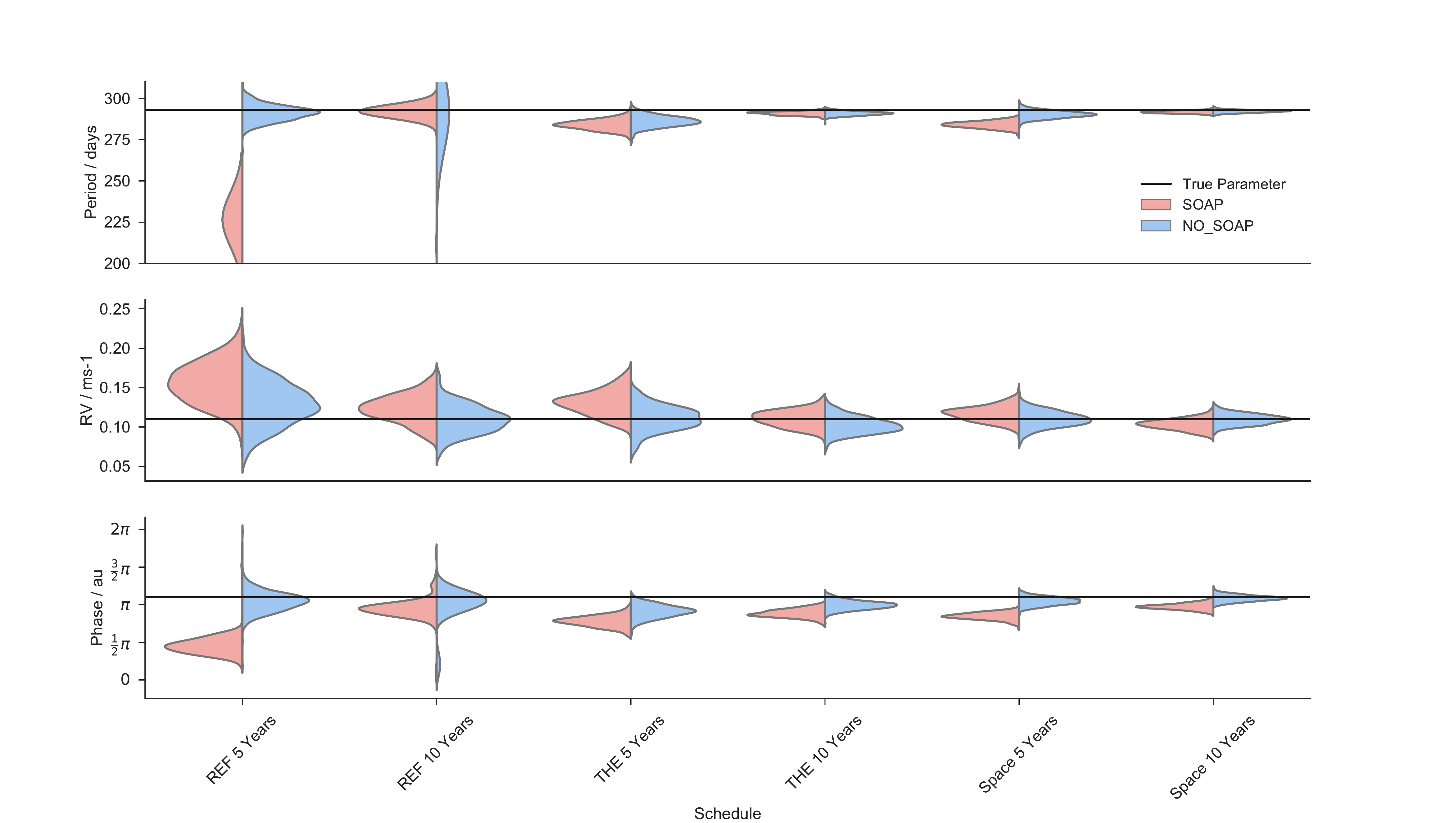}
    \caption{The violin plots of the posterior distributions for the parameters as estimated across all observation schedules for System 1. Each of the $1\times3$ subplots contain the posterior distributions for that parameter as estimated by \PolyChord{} constrained to the correct number of planets and paired for each observation schedule. The red, left--side distributions of each pair are when the data contains the SOAP RVs and Gaussian noise, the blue right--side distributions of each pair are when the data contains just Gaussian noise.}
    \label{fig:sys1_constrained_violin}
\end{figure*}

\begin{figure*}
    \includegraphics[width=1.05\textwidth]{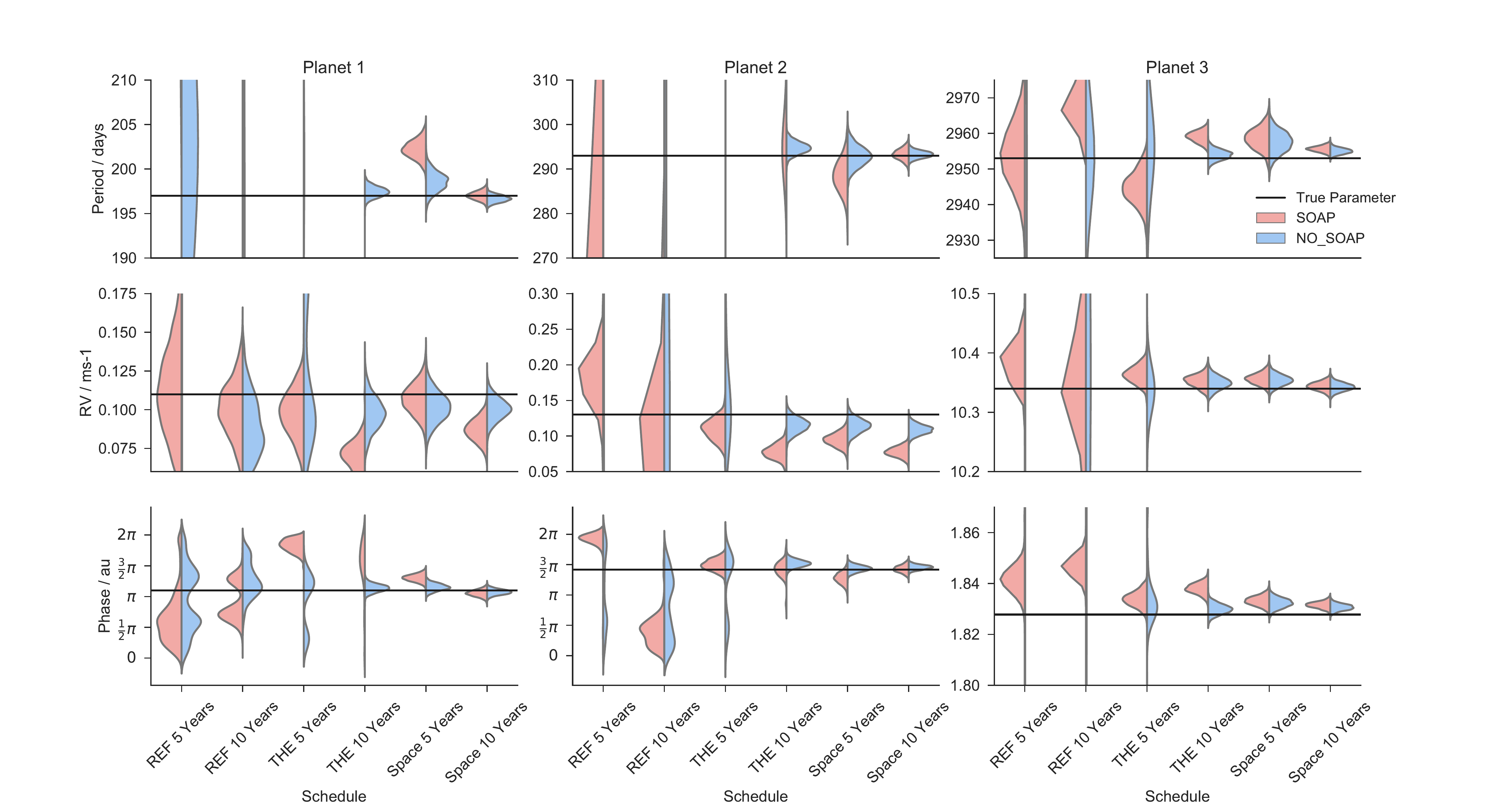}
    \caption{As per Fig. \ref{fig:sys1_constrained_violin} but for System 2, where the model is constrained to the correct 3 planet model.}
    \label{fig:sys2_constrained_violin}
\end{figure*}

\begin{figure*}
    \includegraphics[width=1.05\textwidth]{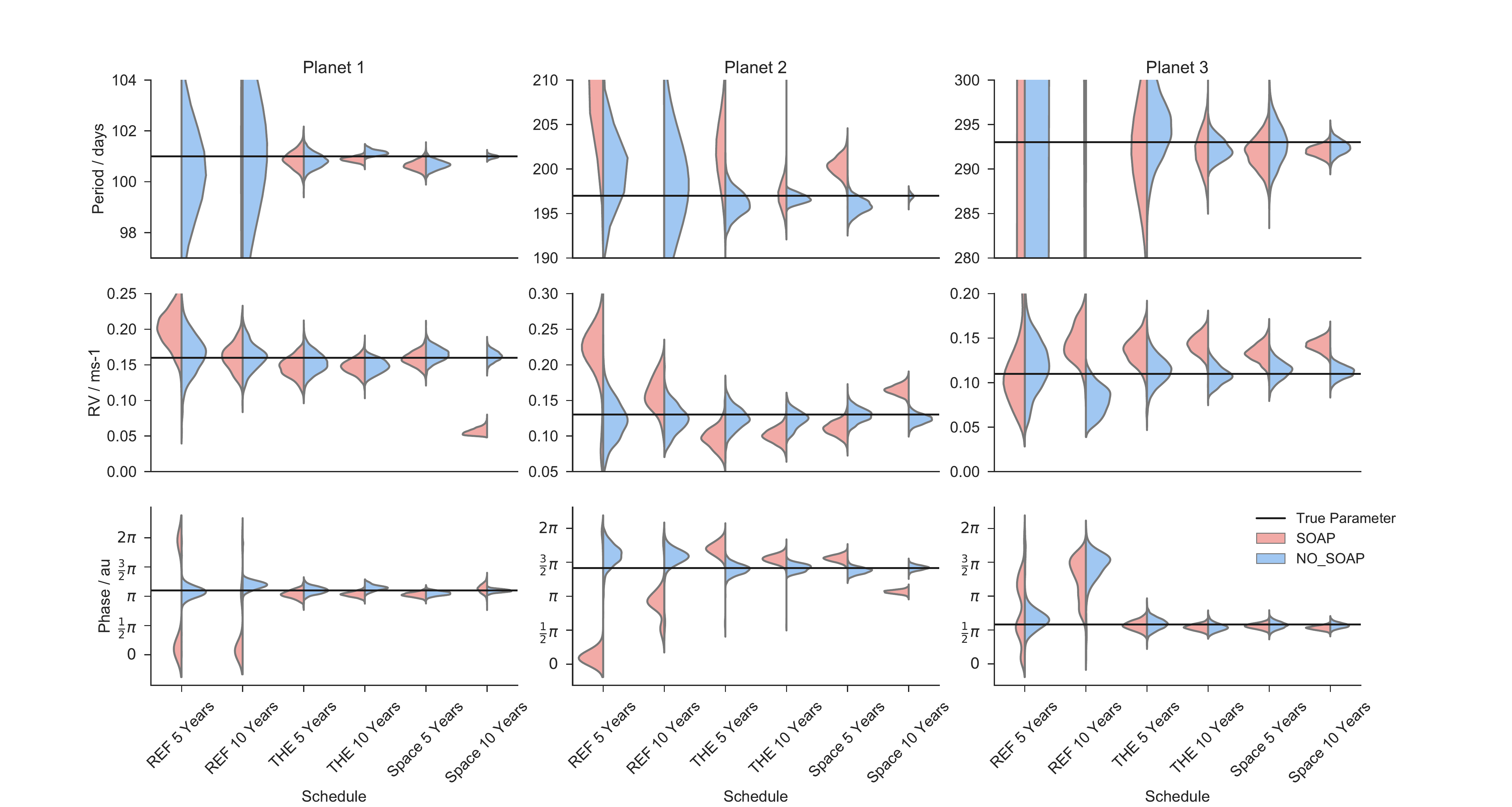}
    \caption{As per Fig. \ref{fig:sys1_constrained_violin} but for System 3, where the model is constrained to the correct 3 planet model.}
    \label{fig:sys3_constrained_violin}
\end{figure*}

To visualise the posterior distributions of the parameter estimates we have grouped the results from section \ref{sec:gauss_results} and section \ref{sec:constrained_results} into violin plots, see Figures \ref{fig:sys1_constrained_violin}, \ref{fig:sys2_constrained_violin}, and \ref{fig:sys3_constrained_violin}, for system 1, 2, and 3 respectively. For a given parameter, e.g. RV semi--amplitude of Planet 1, we show 6 pairs of posteriors with one pair for each observation schedule. The pair contains the posterior of that parameter, with and without the SOAP RVs. This visualisation not only puts the posteriors into context with each other but also allows for a visual investigation into the `quality' of their shape. Generally, narrow Gaussian posteriors are preferred as they represent a normal distribution of solutions with a clear mean and standard deviation, whilst wide and multi-modal distributions indicate that there the results are uncertain and that there are multiple solutions to the parameter estimates. We have clipped the scales of some of the plots to show detail in the narrowest distributions of each set. 

For system 1 in Fig. \ref{fig:sys1_constrained_violin} we can see how the inclusion of SOAP RVs offsets the posterior mean from the true value for most cases and skews the distributions from a smooth Gaussian; however, as you increase the number of observations that effect is lessened. 

For system 2 in Fig. \ref{fig:sys2_constrained_violin}, the SOAP RVs have a much more obvious effect on the posteriors and create many multi--modal distributions for the phase parameter. The very broad widths of the reference schedule posteriors are put into context here with the mostly narrower results from the Terra Hunting and space schedules. Most often, the lowest quality posteriors (multi--modal, non-Gaussian) are associated with the reference schedules. However this quality improves with the improving sampling.

\begin{figure*}
    \includegraphics[width=.95\textwidth]{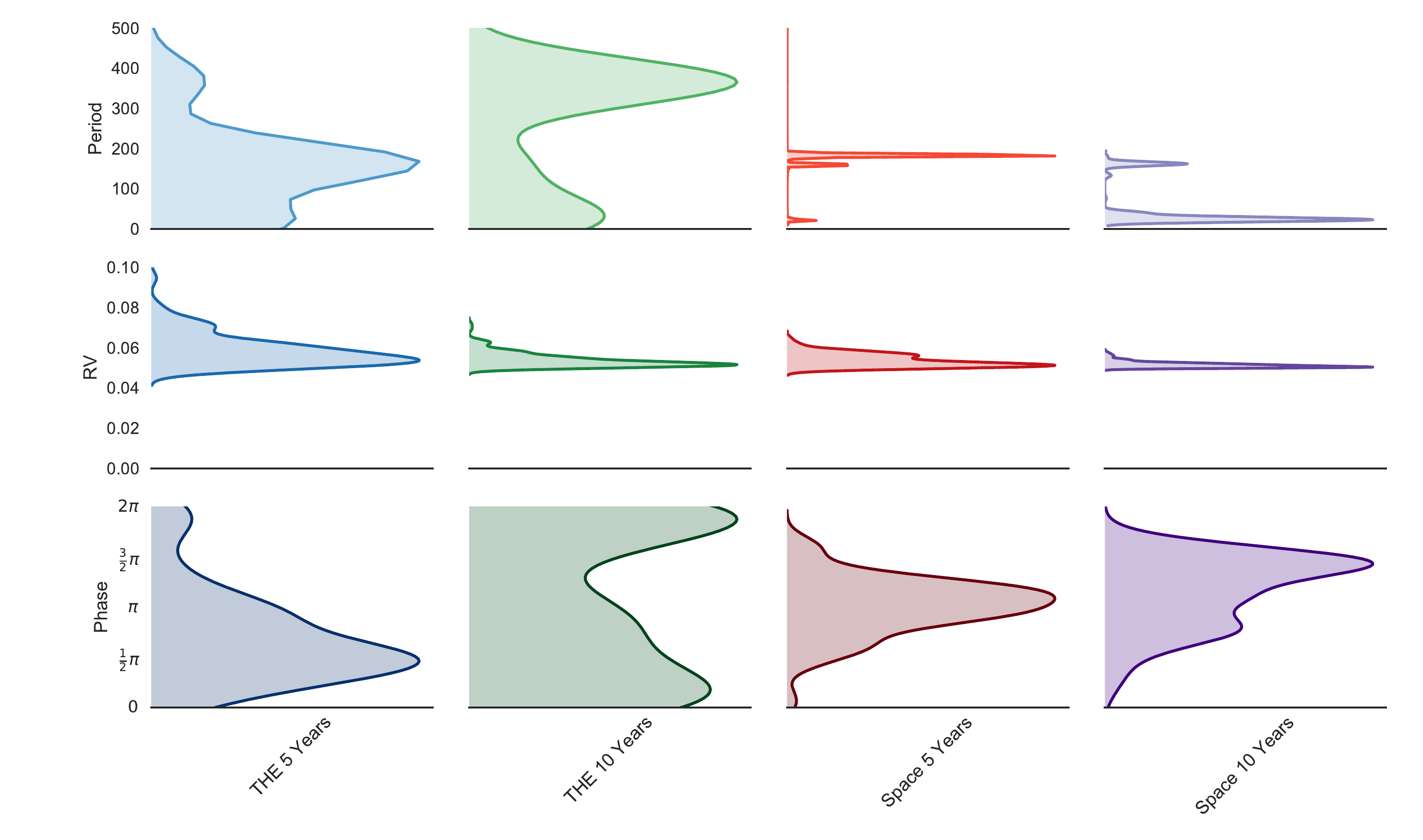}
    \caption{The violin plots of the posterior distributions for System 4 (null planet case) where the data contains no stellar noise. Here, the reference schedules favoured the true 0--planet model so have no posteriors, and we show the falsely favoured 1--planet posteriors of the other four schedules.}
    \label{fig:sys4_nosoap_violin}
\end{figure*}

For system 3 in Fig. \ref{fig:sys3_constrained_violin}, there is again a marked difference between the reference schedules and the others in terms of modality, skewness, and width of the posteriors. As mentioned, in all violin plots some of the posteriors have been clipped due to their large widths often orders of magnitudes larger than others. The last violin plot we show is of the results of the schedules that falsely favoured a 1--planet model for System 4. Figure \ref{fig:sys4_nosoap_violin} shows the 1--planet parameter posteriors for these schedules. Here we can see that \PolyChord{} has fitted very broad and multimodal solutions in period and phase, and has put these solutions at the lower end of our RV prior and so has essentially fitted a very unconstrained signal to the Gaussian noise.

\begin{figure*}
  \includegraphics[width=\textwidth]{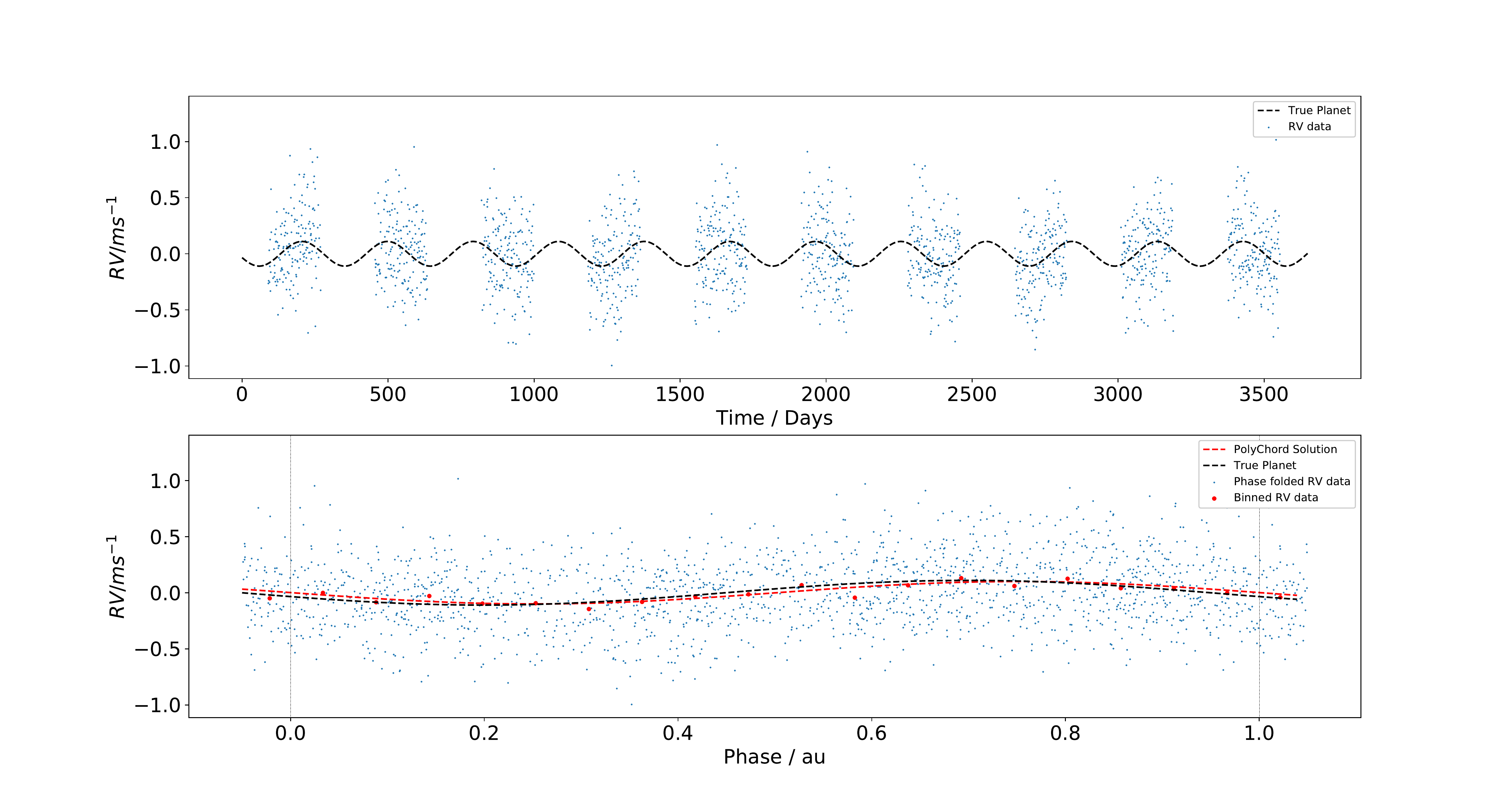}
    \caption{The original (top) and phase--folded (bottom) time series of THE\_$10$ observing System 1 with only Gaussian noise. The solution favoured by \PolyChord{} is drawn on the phase--folded plot as the red dotted line whilst the true planet is the black dotted line.}
    \label{fig:sys1_folded_nosoap}
\end{figure*}

\begin{figure*}
  \includegraphics[width=\textwidth]{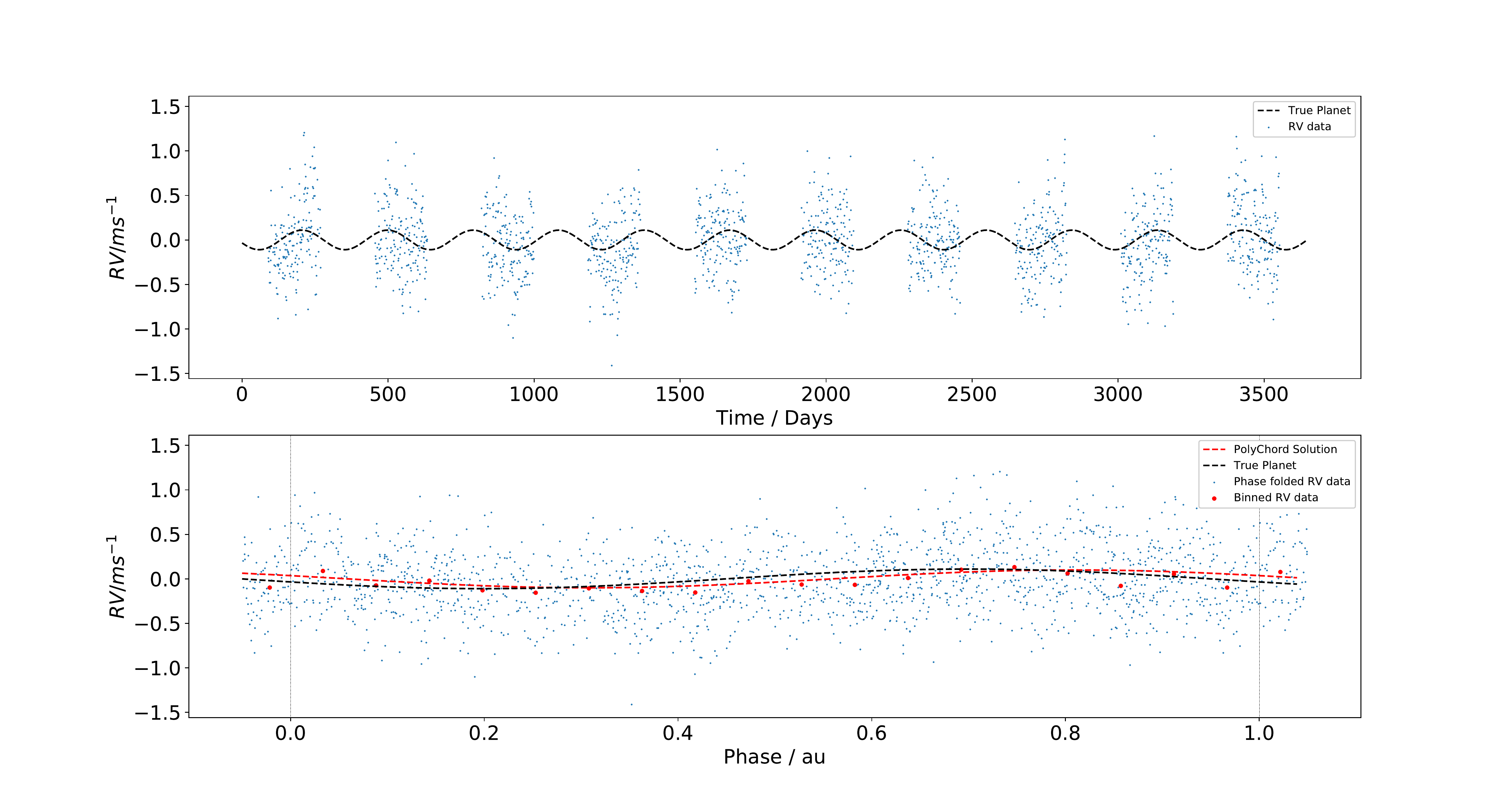}
    \caption{As per Fig. \ref{fig:sys1_folded_nosoap} but with THE\_$10$ observing System 1 with both Gaussian and SOAP noise.}
    \label{fig:sys1_folded_soap}
\end{figure*}

To put the solutions in the context of the original data we fold the original time series, system 1 with and without the SOAP RVs, on the favoured solution of \PolyChord{} and draw the it on top of the binned data to visually show the fit; see Fig. \ref{fig:sys1_folded_nosoap} and \ref{fig:sys1_folded_soap}. In these figures the presence of stellar noise has clearly swayed the \PolyChord{} solution as there is a noticeable phase difference between the recovered solutions when the stellar noise term is included, but not when there is only Gaussian noise. However it does show that it is possible to recover a 300--day, \SI{0.1}{\meter\per\second} signal with the stellar signals present.

\section{Unconstrained model Results from data with Stellar Noise, Gaussian noise, and planets}
\label{sec:unconstrained_results}

We now repeat the analysis on the same data as in section \ref{sec:constrained_results}, but allow \PolyChord{} to assess all planetary models from $0$ $\leq$ $N_p$ $\leq$ $4$. Here we can compare the relative likelihoods of different planetary models and see if our analysis can still find the true planets. We are interested to see where false models are favoured and if we can identify the cause e.g. stellar activity signals, window functions, and aliasing.

\subsection{System 1 Results}
\label{subsec:sys1_soap_unforced}
 
The reference schedules both strongly favoured a $4$ planet model, with the previous parameters of the forced $1$ planet solutions no longer present and all proposed parameters being incorrect. REF\_$5$ finds two candidates at $25.63\pm16.65$ and $46.30\pm51.69$ days which could be an attempt to model the stellar rotation rate as a planet, and it also finds two longer period planets with one being $3399.24\pm500.80$ days at \SI{0.11}{\meter\per\second}, which appears to be another spurious fit, possibly an alias of the baseline. REF\_$10$ also finds a candidate based on the stellar rotation rate but at a larger harmonic, and has three other long--period, low--amplitude planets.

The Terra Hunting schedules fair a little better. THE\_$10$ is surprisingly beaten by THE\_$5$ as it failed to find any correct parameters of the sole Earth--twin. It is possible that it suffered a similar fate to Space\_$10$ versus Space\_$5$ in the previous system 3 results in that we have well--sampled data that has sufficiently resolved the stellar signals so that they become a source of confusion in the analysis due to our lack of modelling the red noise. The THE\_$5$ schedule finds the Earth--twin with three false companions, two of which are stellar rotation based and the last is at $393.91\pm14.33$--days -- likely a window function of our $365$--day basis.

Both space schedules found the Earth-twin along with three companions. For Space\_$5$ the false--positives are all between $20-30$ days and can be attributed to the stellar rotation rate, for Space\_$10$ two of them fall into this category but the third is placed at $109.25\pm0.25$ days. This could either be a harmonic of the stellar rotation, the quasi--periodic $100$--day spot lifetime as dictated by SOAP$2.0$, or that it originates from other window function effects. The full results are shown in Table \ref{tab:sys1_unforced_results}.

\subsection{System 2 Results}
\label{subsec:sys2_soap_unforced}

The reference schedules favoured a $3$ and $4$ planet model for the $5$ and $10$ year baselines respectively, but the evidences for each are marginal to the $4$ and $3$ planet models respectively. REF\_$5$ found the Jupiter analogue, but failed to find any of the low--mass companions. The periods of the false positives also do not match obvious harmonics of the stellar signals but one of them could be a baseline alias as it falls near the $180$ day repeating cycle of observations. REF\_$10$ found  the Jupiter analogue and it appears to have found the Venus analogue but a large error of $198.67\pm175.24$ days, nearly 100\%, which gives low confidence in this detection. The third planet found by this schedule is likely from the stellar rotation.

The Terra Hunting schedules both favoured a $4$ planet model and both included the Jupiter analogue and Earth--twin planets. THE\_$5$ was then accompanied by two low period planets around the stellar rotation periods, whilst THE\_$10$ found a $173.14\pm1.57$ day planet near the $180$ day observation window (the low error suggests it's not a $197$ day Venus candidate) and its fourth planet at $55.14\pm31.41$ which again could be a harmonic of the stellar rotation period.

The Space schedules also strongly favoured a $4$ planet model, but both successfully estimated the parameters of all three planets, and the additional candidate sits at a stellar rotation period. 

It is clear that the reference schedules do not have adequate data density or quantity to fully sample both the stellar signals and the planetary signals, hence our analysis and likelihood function is essentially fitting degenerate solutions to a poorly measured signal. By increasing the number of data points and better distributing the sampling, the Terra Hunting schedules perform much better, to the point of being able to detect the Earth--twin even with our erroneous likelihood function not accounting for the red noise component of this dataset. The space schedules also support this claim, as they have successfully found the signals of all three planets. The full results are shown in Table \ref{tab:sys2_unforced_results}.

\subsection{System 3 Results}
\label{subsec:sys3_soap_unforced}

Both reference schedules strongly favoured a $4$ planet model and find reasonable estimates for the Earth--twin parameters but with large error estimates. For both schedules the companion false positives are a combination of stellar rotation periods and observation schedule windows.

The Terra Hunting schedules performed very well. Despite both strongly favouring a $4$ planet system, they both find good estimates for the three Earth--mass planets and a companion that orbits on the period of the stellar rotation rate. For example the error on the Earth--twin candidates period between THE\_10 and REF\_10 are a factor of 120 smaller.

The space schedules performed comparably to the Terra Hunting schedules, all favouring $4$ planets. Space\_$5$ found the three planets but had large errors on all parameters. The extra planet is a $54.62\pm38.23$--day candidate that is possibly stellar activity related. Space\_$10$ found the three planets with similar estimates to THE\_$10$ and has a false positive at the stellar rotation period again. 

For these results is is clear again that the reference schedules do not have enough coverage or density to detect these planets with confidence when stellar signals are also present. Increasing the quantity of data to that of a Terra Hunting and space schedules is enough for our model to find the planets with confidence even with a likelihood function that does not account for the stellar signals. The full results are shown in Table \ref{tab:sys3_unforced_results}.

\subsection{System 4 Results}
\label{subsec:sys4_soap_unforced}

These data sets only contain the SOAP$2.0$ data and Gaussian noise and are sampled as per the observation schedules. It is therefore a test of how the observing schedule can affect our ability to identify the null--case hypothesis. 

Interestingly, both of the reference schedules were the only ones to favour the zero planet models with significant evidence. This could be that the lack of sampling across all scales simply means that there is no reliable evidence to identify any type of signal, even the stellar signal signals are undersampled.

The Terra Hunting schedules found $3$ and $4$ planets for the $5$ and $10$ year schedules respectively. However each found $2$ planets at low--amplitudes and periods at the stellar rotation rate. Also THE\_$5$ placed a candidate at $379.34\pm39.94$ days which is very close to the $365$ day baseline. THE\_$10$ identified two extra planets at long periods with large errors, both at the lower limit of the RV amplitude prior suggesting that the solution did not converge.

The Space\_$5$ schedule found $4$ planets all at long periods that could be aliases of the exact $5$--year window of the measurement series. Space\_$10$ found three planets, two at the stellar rotation rate and a third up at $891.00\pm15.61$ days. We look at the window function in more detail in section \ref{sec:FAP_future}.

The need for some degree of stellar modelling is most apparent here to identify the key activity cycles within the data. Whilst the denser series can partly overcome this with enough data, they are still susceptible to these quasi--periodic signals. We limited our search to $4$ planets for reasons of computational expense, but it is possible that \PolyChord{} would fit an ever increasing number of planets to the stellar noise if left completely unconstrained. However it is interesting that even with our inadequate modelling, we are still able to recover a \SI{0.10}{\meter\per\second} signal with and without stellar noise, and with and without planetary companions with enough sampling in some cases. The full results are shown in Table \ref{tab:sys4_unforced_results}.

\begin{figure*}
    \includegraphics[width=\textwidth]{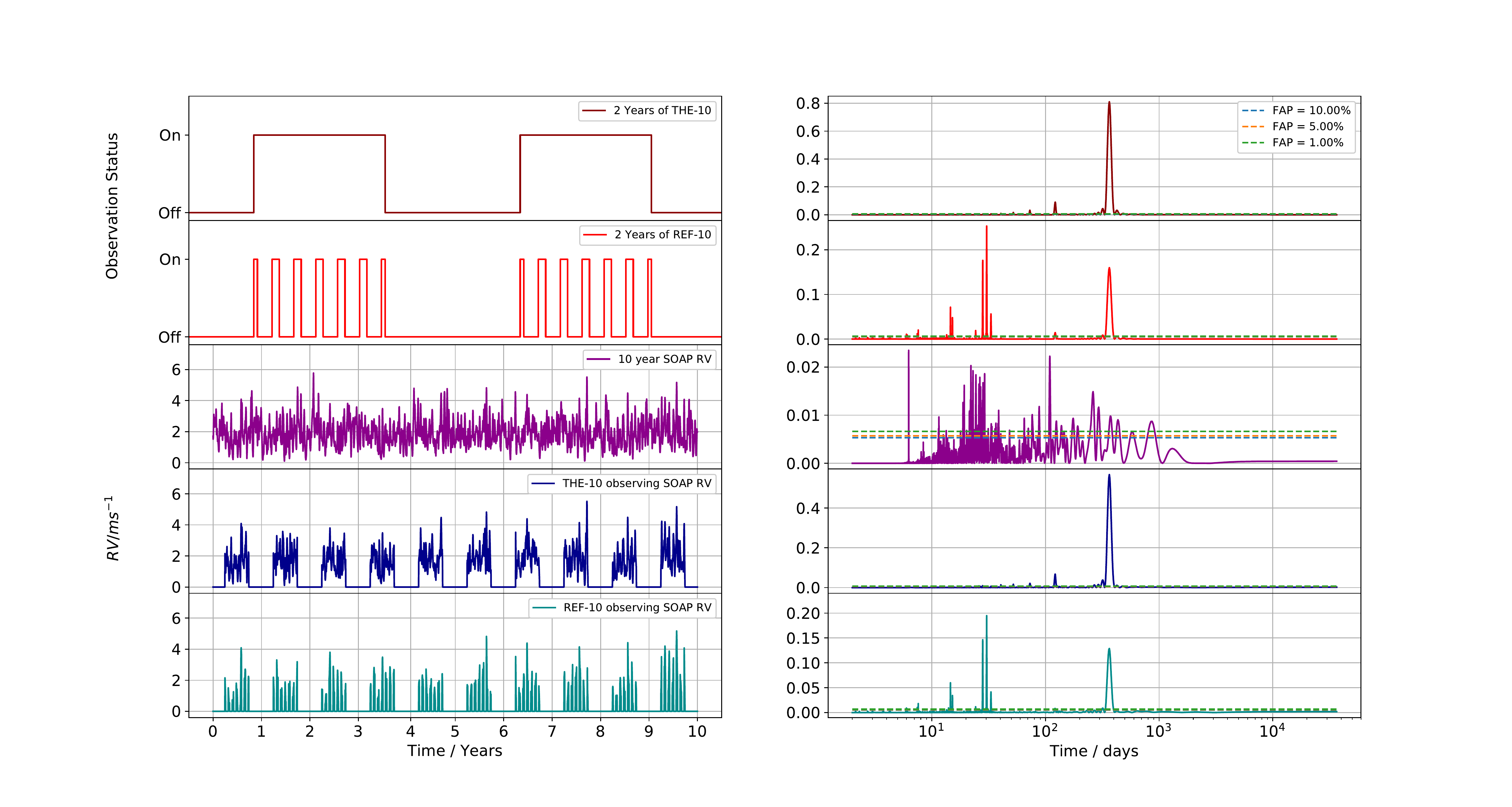}
    \centering
    \caption{An example of window function analysis for two observation schedules observing stellar signals from a quiet star. From top to bottom, each pair of plots (left and right in the same row, same colour) are: a 2--year zoom of Terra Hunting Experiment schedule and the Lomb--Scargle (LS) of that time series, a 2--year zoom of the reference schedule and the corresponding LS, a $10$--year SOAP generated RV for a quiet star and the corresponding LS, the THE--$10$ schedule observing the stellar signals and the corresponding LS, and finally the REF--$10$ schedule observing the stellar signals and the corresponding LS. On each periodogram we plot the $10$\%, $5$\%, and $1$\% false--alarm probabilities }
    \label{fig:window_func}
\end{figure*}


\section{False Alarms, Results summary, And future work}
\label{sec:FAP_future}
Both the SOAP RVs and our observation schedule have introduced periodicities into the data which may be interpreted as planets and which explain some of the false periods favoured by \PolyChord{} at various points in the analysis in some cases. To make contact with a traditional approach commonly seen in the community, we perform a simple frequency analysis to identify periodicities in the data by taking the System 4 data (no planets) observed by each observation schedule, and producing a Lomb-Scargle periodogram of the time series. Any significant peaks will be a result of the observation windows and the periodic signals present in the SOAP data. Figure~\ref{fig:window_func} displays the observation schedule status of THE--$10$ and REF--$10$ as a binary plot (a two year zoom, and omitting weather downtime) and the full stellar RV series from SOAP (also the same sampling as the Space\_$10$ schedule). We then combine these to obtain an RV series of  as seen by each schedule, which is essentially the unattenuated original SOAP data. A Lomb-Scargle periodogram is produced at each step to see the various periods that may be present as false positives.

By plotting various false--alarm probabilities we can see that for each schedule there are many significant peaks above the $1$\% threshold. The THE--$10$ schedule shows an expected significant peak at $365$ days along with minor harmonics at $120$, $73$ and $50$ days. The REF--$10$ schedule has more significant peaks at around $30$ days due to the monthly cadence of observations. The periodogram of the SOAP$2.0$ data shows multiple large peaks between $20-40$ days, which correspond to the stellar rotation period, and a large peak at $100$ days; the average spot group lifetime. We have found many instances of false--positives at or near these periods across all observation schedules analyses. Even with excellent sampling in the face of stellar noise, we would be cautious about accepting any low amplitude planets at these periods without a more complete analysis.

To combine all of the results into a single metric, we have created a scoring system to rank the observation schedules performance against each other. There are three sets of results to rank: the case with white noise but no stellar noise, including the stellar noise but constraining the models, and including stellar noise and leaving the analysis unconstrained. For each set, we have created a decision tree that allocates points to each observation schedule based off of its ability to chose the correct model and find true planetary signals. 

\begin{figure}
    \includegraphics[width=\columnwidth]{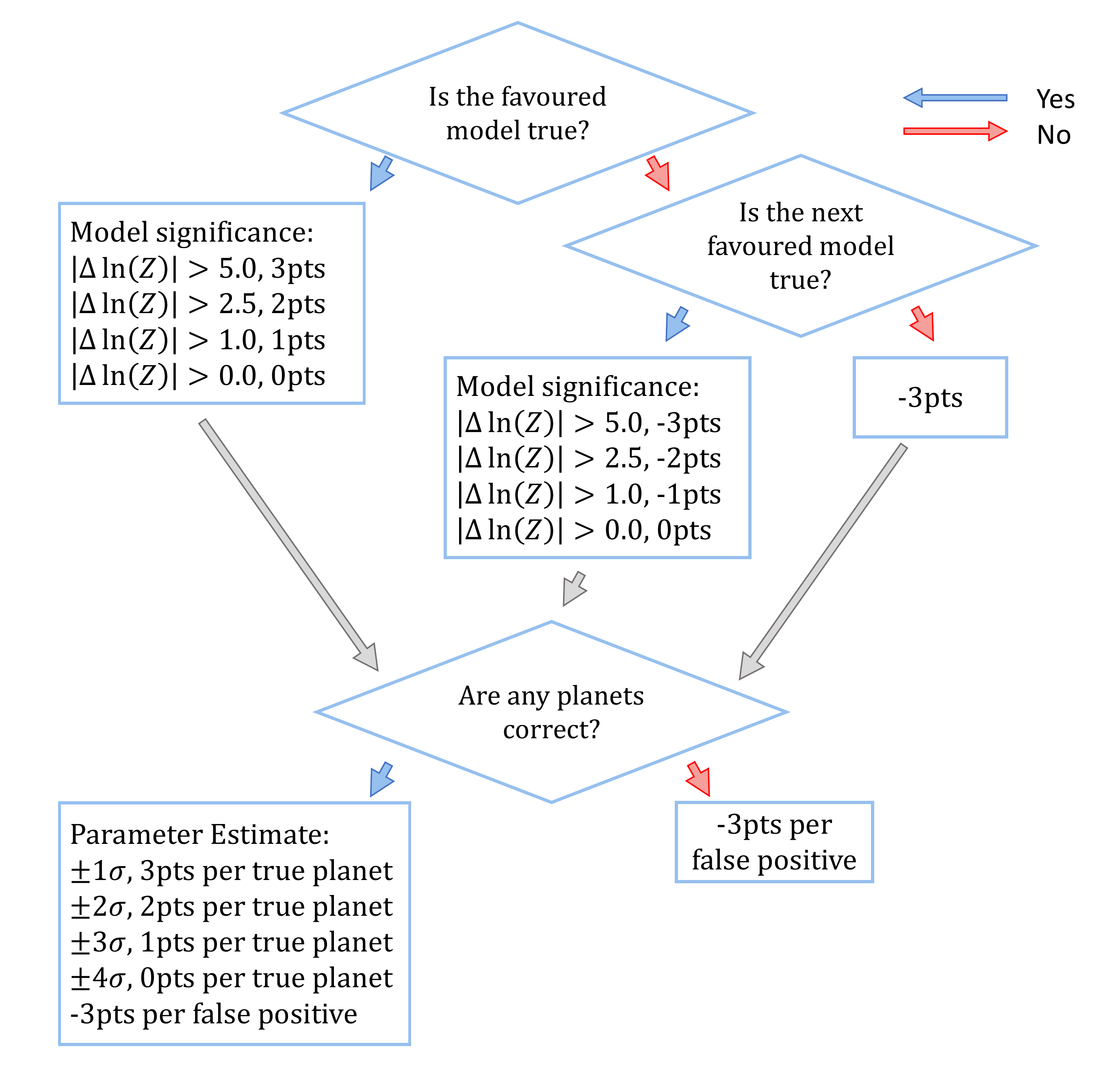}
    \centering
    \caption{The scoring system and decision tree used to rank the observation schedules.}
    \label{fig:flow_chart}
\end{figure}

\begin{figure*}
    \includegraphics[width=\textwidth]{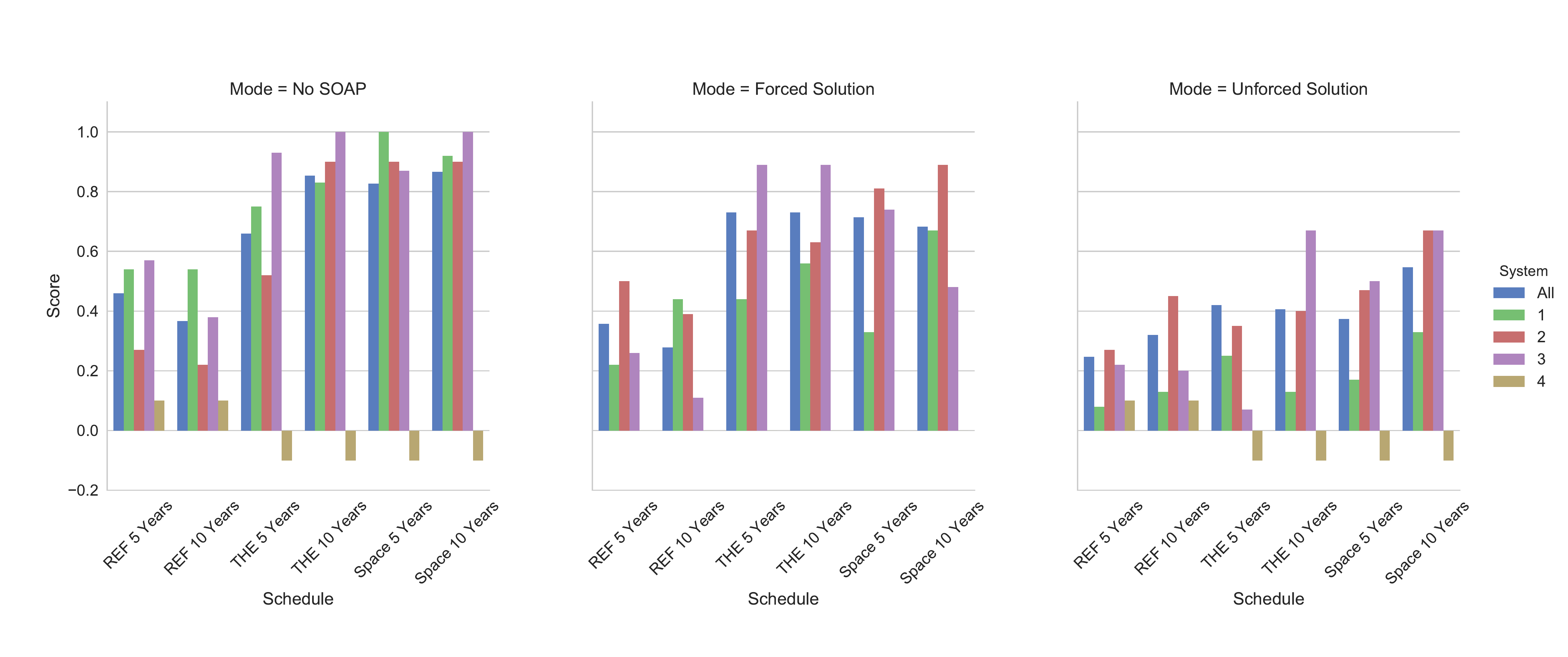}
    \centering
    \caption{The results of scoring the models as per the flow chart in Fig. \ref{fig:flow_chart}. The scores are normalised to the maximum possible score of each solar system, grouped into observation schedules, and plotted as an array of each mode of the analysis. The first bar in each group shows the observation schedule's performance across all solar systems, denoted by `all' which is an average of the scores, then each individual score is shown after in order from system 1 to system 4.}
    \label{fig:score}
\end{figure*}

The scoring system will also penalise a model for strongly favouring a false model or false--positives. Figure \ref{fig:flow_chart} shows our scoring system and the decision tree, and in Figure \ref{fig:score} we see the overall performance for each schedule and the related analysis mode: without stellar noise, forced solutions, or unconstrained solutions. Here we can see that the performance generally improves for a denser schedule. In almost all cases the reference schedules performed poorly, rarely achieving a score above $30\%$ of maximum, whilst the Terra Hunting and space schedules performed well, both regularly scoring $60-70\%$ of maximum, and over $80\%$ where we have proper noise modelling. It is interesting to note that despite the almost two--fold increase of data available to the space schedules, its performance is only marginally better than Terra Hunting. With perfect noise modelling and $N$ measurements, we expect the performance to increase as $1/\sqrt N$, however we do not observe this. This is very likely due to the incorrect stellar noise model becoming a hindrance or barrier that the schedules can not overcome with better sampling alone. This hindrance becomes apparent when we have allowed \PolyChord{} free choice of the model, as shown in the $3$rd plot, as the performance has dropped across all schedules but maintained the general trend. Also interesting are the scores for System $4$, only the reference schedule managed to successfully identify the zero--planet model whilst both the Terra Hunting and Space schedules always favoured solutions with one or more planets. However, these false--positives were often at periods likely introduced by SOAP. We hope that with adequate stellar modelling the performance of the unconstrained mode could be boosted to match the constrained data set. This will be investigated with follow--up work.

We acknowledge the assumptions we have made regarding the outcome of stellar modelling to be overly simplistic. In follow--up work we aim to include a more realistic scheduling using the Terra Hunting Scheduler which is currently under development and we will factor in stellar variability into our data analysis by means of including short--term quasi--periodic noise modelling into our likelihood function. We also aim to increase the number of observed systems to be representative of a true yield estimation simulation by generating many hundreds of stable systems, using observed planet occurrence rates and different stellar signals, and then use a representative observation schedule from the Terra Hunting Experiment schedule (in prep). 

\PolyChord{} operates on an intrinsically random process, hence we have run \PolyChord{} on each data set three times to check the robustness and reliability of our results. The slight randomness of the nested sampling could lead us to accept a local maximum instead of the global maximum. In all three of our \PolyChord{} runs we found the results to be consistent within error, largely due to us using a sufficiently high number of live points to explore the posterior probability space.


\section{Conclusions}

We have presented our work on simulating results for the Terra Hunting Experiment in comparison to current typical RV surveys. The multi-nested analysis has proven to be both a powerful and robust tool for fitting a simple model to complex data.

We have demonstrated that the proposed schedule of the Terra Hunting experiment typically out-performs a standard RV survey in three different solar system architectures. The Terra Hunting and Space schedules found many false--positives that arise from the unmodelled stellar signals which is also promising from the prospect of using such an instrument for the explicit purpose of observing stellar signals. The proposed schedule of the Terra Hunting Experiment performed comparably to a space--based observation schedule in its retrieval of $1M_{\earth}$ planets in the habitable zone of Sun--like stars.

In our analysis we only accounted for random white noise however we have demonstrated the improvement in planetary parameter estimates that can be gained by using better sampling strategies, even for those datasets that contained red noise arising from stellar activity.

We plan to use this study as a foundation for a more detailed investigation into RV sampling for the application of Earth--twin discovery, and for the Terra Hunting Experiment. Follow--up work to further refine the simulation and analysis pipeline is under way.

\section*{Acknowledgements}

The authors thank X. Dumusque in multiple troubleshooting occasions of installing and running SOAP2.0. We thank G. Willatt for his extensive support of the Cavendish Astrophysics Cluster on which nearly all of the results were obtained. We also thank V. Hod\v{z}i\'c for his assistance in the script to generate planetary radial velocities, J. Briegal for his phase--folding plotting tool, and D. Green for his help in formatting the \LaTeX{} for the paper.

The authors also thank the anonymous reviewer for their many detailed and helpful comments.

S. J. Thompson and D. Queloz acknowledges the support from the Science and Technologies Facilities Council (STFC) as part of research grant ST/N002997/1.  R. Hall acknowledges the STFC for his PhD studentship award number 1641620, and M. Hobson for discussions on multi-nested sampling. W. Handley acknowledges online discussions regarding switching degeneracy between sets of planet parameters \citep{AstroStatistics}.




\clearpage
\bibliographystyle{mnras}
\bibliography{bibliography} 

\begin{thebibliography}{}
\makeatletter
\relax
\def\mn@urlcharsother{\let\do\@makeother \do\$\do\&\do\#\do\^\do\_\do\%\do\~}
\def\mn@doi{\begingroup\mn@urlcharsother \@ifnextchar [ {\mn@doi@}
  {\mn@doi@[]}}
\def\mn@doi@[#1]#2{\def\@tempa{#1}\ifx\@tempa\@empty \href
  {http://dx.doi.org/#2} {doi:#2}\else \href {http://dx.doi.org/#2} {#1}\fi
  \endgroup}
\def\mn@eprint#1#2{\mn@eprint@#1:#2::\@nil}
\def\mn@eprint@arXiv#1{\href {http://arxiv.org/abs/#1} {{\tt arXiv:#1}}}
\def\mn@eprint@dblp#1{\href {http://dblp.uni-trier.de/rec/bibtex/#1.xml}
  {dblp:#1}}
\def\mn@eprint@#1:#2:#3:#4\@nil{\def\@tempa {#1}\def\@tempb {#2}\def\@tempc
  {#3}\ifx \@tempc \@empty \let \@tempc \@tempb \let \@tempb \@tempa \fi \ifx
  \@tempb \@empty \def\@tempb {arXiv}\fi \@ifundefined
  {mn@eprint@\@tempb}{\@tempb:\@tempc}{\expandafter \expandafter \csname
  mn@eprint@\@tempb\endcsname \expandafter{\@tempc}}}

\bibitem[\protect\citeauthoryear{AstroStatistics}{AstroStatistics}{2013}]{AstroStatistics}
AstroStatistics 2013, Part 2: Forced identifiability for exoplanets, \url
  {https://astrostatistics.wordpress.com/2013/08/06/part-2-forced-identifiability-for-exoplanets/}

\bibitem[\protect\citeauthoryear{{Baranne} et~al.,}{{Baranne}
  et~al.}{1996}]{1996A&AS..119..373B}
{Baranne} A.,  et~al., 1996, \aaps, \href
  {http://adsabs.harvard.edu/abs/1996A%26AS..119..373B} {119, 373}

\bibitem[\protect\citeauthoryear{{Borucki} et~al.,}{{Borucki}
  et~al.}{2010}]{2010Sci...327..977B}
{Borucki} W.~J.,  et~al., 2010, \mn@doi [Science] {10.1126/science.1185402},
  \href {http://adsabs.harvard.edu/abs/2010Sci...327..977B} {327, 977}

\bibitem[\protect\citeauthoryear{Clubb}{Clubb}{2008}]{web_RV}
Clubb K.,  2008, A Detailed Derivation of The Radial Velocity Equation, \url
  {http://www.relativitycalculator.com/pdfs/RV_Derivation.pdf}

\bibitem[\protect\citeauthoryear{{Cosentino} et~al.,}{{Cosentino}
  et~al.}{2012}]{2012SPIE.8446E..1VC}
{Cosentino} R.,  et~al., 2012, in Ground-based and Airborne Instrumentation for
  Astronomy IV. p. 84461V

\bibitem[\protect\citeauthoryear{{Dumusque}}{{Dumusque}}{2016}]{2016A&A...593A...5D}
{Dumusque} X.,  2016, \mn@doi [\aap] {10.1051/0004-6361/201628672}, \href
  {http://adsabs.harvard.edu/abs/2016A%26A...593A...5D} {593, A5}

\bibitem[\protect\citeauthoryear{{Dumusque}, {Udry}, {Lovis}, {Santos}  \&
  {Monteiro}}{{Dumusque} et~al.}{2011}]{2011A&A...525A.140D}
{Dumusque} X.,  {Udry} S.,  {Lovis} C.,  {Santos} N.~C.,   {Monteiro}
  M.~J.~P.~F.~G.,  2011, \mn@doi [\aap] {10.1051/0004-6361/201014097}, \href
  {http://adsabs.harvard.edu/abs/2011A%26A...525A.140D} {525, A140}

\bibitem[\protect\citeauthoryear{{Dumusque} et~al.,}{{Dumusque}
  et~al.}{2012}]{2012Natur.491..207D}
{Dumusque} X.,  et~al., 2012, \mn@doi [\nat] {10.1038/nature11572}, \href
  {http://adsabs.harvard.edu/abs/2012Natur.491..207D} {491, 207}

\bibitem[\protect\citeauthoryear{{Dumusque}, {Boisse}  \& {Santos}}{{Dumusque}
  et~al.}{2014}]{2014ApJ...796..132D}
{Dumusque} X.,  {Boisse} I.,   {Santos} N.~C.,  2014, \mn@doi [\apj]
  {10.1088/0004-637X/796/2/132}, \href
  {http://adsabs.harvard.edu/abs/2014ApJ...796..132D} {796, 132}

\bibitem[\protect\citeauthoryear{{Dumusque}, {Boisse}  \& {Santos}}{{Dumusque}
  et~al.}{2015}]{2015ascl.soft04021D}
{Dumusque} X.,  {Boisse} I.,   {Santos} N.~C.,  2015, {SOAP 2.0: Spot
  Oscillation And Planet 2.0}, Astrophysics Source Code Library (\mn@eprint
  {ascl} {1504.021})

\bibitem[\protect\citeauthoryear{{Dumusque} et~al.,}{{Dumusque}
  et~al.}{2017}]{2017A&A...598A.133D}
{Dumusque} X.,  et~al., 2017, \mn@doi [\aap] {10.1051/0004-6361/201628671},
  \href {http://adsabs.harvard.edu/abs/2017A%26A...598A.133D} {598, A133}

\bibitem[\protect\citeauthoryear{{Feroz} \& {Hobson}}{{Feroz} \&
  {Hobson}}{2008}]{2008MNRAS.384..449F}
{Feroz} F.,  {Hobson} M.~P.,  2008, \mn@doi [\mnras]
  {10.1111/j.1365-2966.2007.12353.x}, \href
  {http://adsabs.harvard.edu/abs/2008MNRAS.384..449F} {384, 449}

\bibitem[\protect\citeauthoryear{{Feroz} \& {Hobson}}{{Feroz} \&
  {Hobson}}{2014}]{2014MNRAS.437.3540F}
{Feroz} F.,  {Hobson} M.~P.,  2014, \mn@doi [\mnras] {10.1093/mnras/stt2148},
  \href {http://adsabs.harvard.edu/abs/2014MNRAS.437.3540F} {437, 3540}

\bibitem[\protect\citeauthoryear{{Feroz}, {Hobson}  \& {Bridges}}{{Feroz}
  et~al.}{2009}]{2009MNRAS.398.1601F}
{Feroz} F.,  {Hobson} M.~P.,   {Bridges} M.,  2009, \mn@doi [\mnras]
  {10.1111/j.1365-2966.2009.14548.x}, \href
  {http://adsabs.harvard.edu/abs/2009MNRAS.398.1601F} {398, 1601}

\bibitem[\protect\citeauthoryear{{Feroz}, {Balan}  \& {Hobson}}{{Feroz}
  et~al.}{2011}]{2011MNRAS.415.3462F}
{Feroz} F.,  {Balan} S.~T.,   {Hobson} M.~P.,  2011, \mn@doi [\mnras]
  {10.1111/j.1365-2966.2011.18962.x}, \href
  {http://adsabs.harvard.edu/abs/2011MNRAS.415.3462F} {415, 3462}

\bibitem[\protect\citeauthoryear{{Feroz}, {Hobson}, {Cameron}  \&
  {Pettitt}}{{Feroz} et~al.}{2013}]{2013arXiv1306.2144F}
{Feroz} F.,  {Hobson} M.~P.,  {Cameron} E.,   {Pettitt} A.~N.,  2013, preprint,
  \href {http://adsabs.harvard.edu/abs/2013arXiv1306.2144F} {} (\mn@eprint
  {arXiv} {1306.2144})

\bibitem[\protect\citeauthoryear{{Fischer} et~al.,}{{Fischer}
  et~al.}{2016}]{2016PASP..128f6001F}
{Fischer} D.~A.,  et~al., 2016, \mn@doi [\pasp]
  {10.1088/1538-3873/128/964/066001}, \href
  {http://adsabs.harvard.edu/abs/2016PASP..128f6001F} {128, 066001}

\bibitem[\protect\citeauthoryear{{Gregory}}{{Gregory}}{2007}]{2007MNRAS.374.1321G}
{Gregory} P.~C.,  2007, \mn@doi [\mnras] {10.1111/j.1365-2966.2006.11240.x},
  \href {http://adsabs.harvard.edu/abs/2007MNRAS.374.1321G} {374, 1321}

\bibitem[\protect\citeauthoryear{Hall, Thompson  \& Queloz}{Hall
  et~al.}{2016}]{harps3_CCD_spie}
Hall R.,  Thompson S.,   Queloz D.,  2016, SPIE Proceedings High Energy,
  Optical, and Infrared Detectors for Astronomy VII, 9915

\bibitem[\protect\citeauthoryear{{Handley}, {Hobson}  \& {Lasenby}}{{Handley}
  et~al.}{2015a}]{2015MNRAS.450L..61H}
{Handley} W.~J.,  {Hobson} M.~P.,   {Lasenby} A.~N.,  2015a, \mn@doi [\mnras]
  {10.1093/mnrasl/slv047}, \href
  {http://adsabs.harvard.edu/abs/2015MNRAS.450L..61H} {450, L61}

\bibitem[\protect\citeauthoryear{{Handley}, {Hobson}  \& {Lasenby}}{{Handley}
  et~al.}{2015b}]{2015MNRAS.453.4384H}
{Handley} W.~J.,  {Hobson} M.~P.,   {Lasenby} A.~N.,  2015b, \mn@doi [\mnras]
  {10.1093/mnras/stv1911}, \href
  {http://adsabs.harvard.edu/abs/2015MNRAS.453.4384H} {453, 4384}

\bibitem[\protect\citeauthoryear{{Haywood} et~al.,}{{Haywood}
  et~al.}{2016}]{2016MNRAS.457.3637H}
{Haywood} R.~D.,  et~al., 2016, \mn@doi [\mnras] {10.1093/mnras/stw187}, \href
  {http://adsabs.harvard.edu/abs/2016MNRAS.457.3637H} {457, 3637}

\bibitem[\protect\citeauthoryear{Jeffreys}{Jeffreys}{1983}]{jeffreys1983theory}
Jeffreys H.,  1983, Theory of Probability.
International series of monographs on physics, Clarendon Press, \url
  {https://books.google.co.uk/books?id=EbodAQAAMAAJ}

\bibitem[\protect\citeauthoryear{{Jurgenson}, {Fischer}, {McCracken}, {Sawyer},
  {Szymkowiak}, {Davis}, {Muller}  \& {Santoro}}{{Jurgenson}
  et~al.}{2016}]{2016SPIE.9908E..6TJ}
{Jurgenson} C.,  {Fischer} D.,  {McCracken} T.,  {Sawyer} D.,  {Szymkowiak} A.,
   {Davis} A.,  {Muller} G.,   {Santoro} F.,  2016, in Ground-based and
  Airborne Instrumentation for Astronomy VI. p. 99086T (\mn@eprint {arXiv}
  {1606.04413})

\bibitem[\protect\citeauthoryear{{Kane} et~al.,}{{Kane}
  et~al.}{2016}]{2016ApJ...830....1K}
{Kane} S.~R.,  et~al., 2016, \mn@doi [\apj] {10.3847/0004-637X/830/1/1}, \href
  {http://adsabs.harvard.edu/abs/2016ApJ...830....1K} {830, 1}

\bibitem[\protect\citeauthoryear{{Lagrange}, {Meunier}, {Desort}  \&
  {Malbet}}{{Lagrange} et~al.}{2011}]{2011A&A...528L...9L}
{Lagrange} A.-M.,  {Meunier} N.,  {Desort} M.,   {Malbet} F.,  2011, \mn@doi
  [\aap] {10.1051/0004-6361/201016354}, \href
  {http://adsabs.harvard.edu/abs/2011A%26A...528L...9L} {528, L9}

\bibitem[\protect\citeauthoryear{MacKay}{MacKay}{2003}]{mackay2003information}
MacKay D.~J.,  2003, Information theory, inference and learning algorithms.
Cambridge university press

\bibitem[\protect\citeauthoryear{Mayor \& Queloz}{Mayor \&
  Queloz}{1995}]{Mayor:1995aa}
Mayor M.,  Queloz D.,  1995, Nature, 378, 355

\bibitem[\protect\citeauthoryear{{Mayor} et~al.,}{{Mayor}
  et~al.}{2003}]{2003Msngr.114...20M}
{Mayor} M.,  et~al., 2003, The Messenger, \href
  {http://adsabs.harvard.edu/abs/2003Msngr.114...20M} {114, 20}

\bibitem[\protect\citeauthoryear{{Meunier} \& {Lagrange}}{{Meunier} \&
  {Lagrange}}{2013}]{2013A&A...551A.101M}
{Meunier} N.,  {Lagrange} A.-M.,  2013, \mn@doi [\aap]
  {10.1051/0004-6361/201219917}, \href
  {http://adsabs.harvard.edu/abs/2013A%26A...551A.101M} {551, A101}

\bibitem[\protect\citeauthoryear{{Noyes}, {Hartmann}, {Baliunas}, {Duncan}  \&
  {Vaughan}}{{Noyes} et~al.}{1984}]{1984ApJ...279..763N}
{Noyes} R.~W.,  {Hartmann} L.~W.,  {Baliunas} S.~L.,  {Duncan} D.~K.,
  {Vaughan} A.~H.,  1984, \mn@doi [\apj] {10.1086/161945}, \href
  {http://adsabs.harvard.edu/abs/1984ApJ...279..763N} {279, 763}

\bibitem[\protect\citeauthoryear{{Pepe} et~al.,}{{Pepe}
  et~al.}{2011}]{2011A&A...534A..58P}
{Pepe} F.,  et~al., 2011, \mn@doi [\aap] {10.1051/0004-6361/201117055}, \href
  {http://adsabs.harvard.edu/abs/2011A%26A...534A..58P} {534, A58}

\bibitem[\protect\citeauthoryear{{Pepe} et~al.,}{{Pepe}
  et~al.}{2014}]{2014arXiv1401.5918P}
{Pepe} F.,  et~al., 2014, preprint, \href
  {http://adsabs.harvard.edu/abs/2014arXiv1401.5918P} {} (\mn@eprint {arXiv}
  {1401.5918})

\bibitem[\protect\citeauthoryear{{Pollacco} et~al.,}{{Pollacco}
  et~al.}{2006}]{2006PASP..118.1407P}
{Pollacco} D.~L.,  et~al., 2006, \mn@doi [\pasp] {10.1086/508556}, \href
  {http://adsabs.harvard.edu/abs/2006PASP..118.1407P} {118, 1407}

\bibitem[\protect\citeauthoryear{{Queloz} et~al.,}{{Queloz}
  et~al.}{2001}]{2001A&A...379..279Q}
{Queloz} D.,  et~al., 2001, \mn@doi [\aap] {10.1051/0004-6361:20011308}, \href
  {http://adsabs.harvard.edu/abs/2001A%26A...379..279Q} {379, 279}

\bibitem[\protect\citeauthoryear{{Rajpaul}, {Aigrain}, {Osborne}, {Reece}  \&
  {Roberts}}{{Rajpaul} et~al.}{2015}]{2015MNRAS.452.2269R}
{Rajpaul} V.,  {Aigrain} S.,  {Osborne} M.~A.,  {Reece} S.,   {Roberts} S.,
  2015, \mn@doi [\mnras] {10.1093/mnras/stv1428}, \href
  {http://adsabs.harvard.edu/abs/2015MNRAS.452.2269R} {452, 2269}

\bibitem[\protect\citeauthoryear{{Rajpaul}, {Aigrain}  \& {Roberts}}{{Rajpaul}
  et~al.}{2016}]{2016MNRAS.456L...6R}
{Rajpaul} V.,  {Aigrain} S.,   {Roberts} S.,  2016, \mn@doi [\mnras]
  {10.1093/mnrasl/slv164}, \href
  {http://adsabs.harvard.edu/abs/2016MNRAS.456L...6R} {456, L6}

\bibitem[\protect\citeauthoryear{{Rajpaul}, {Buchhave}  \& {Aigrain}}{{Rajpaul}
  et~al.}{2017}]{2017MNRAS.471L.125R}
{Rajpaul} V.,  {Buchhave} L.~A.,   {Aigrain} S.,  2017, \mn@doi [\mnras]
  {10.1093/mnrasl/slx116}, \href
  {http://adsabs.harvard.edu/abs/2017MNRAS.471L.125R} {471, L125}

\bibitem[\protect\citeauthoryear{{Ricker} et~al.,}{{Ricker}
  et~al.}{2014}]{2014SPIE.9143E..20R}
{Ricker} G.~R.,  et~al., 2014, in Space Telescopes and Instrumentation 2014:
  Optical, Infrared, and Millimeter Wave. p. 914320

\bibitem[\protect\citeauthoryear{{Shapiro}, {Solanki}, {Krivova}, {Schmutz},
  {Ball}, {Knaack}, {Rozanov}  \& {Unruh}}{{Shapiro}
  et~al.}{2014}]{2014A&A...569A..38S}
{Shapiro} A.~I.,  {Solanki} S.~K.,  {Krivova} N.~A.,  {Schmutz} W.~K.,  {Ball}
  W.~T.,  {Knaack} R.,  {Rozanov} E.~V.,   {Unruh} Y.~C.,  2014, \mn@doi [\aap]
  {10.1051/0004-6361/201323086}, \href
  {http://adsabs.harvard.edu/abs/2014A%26A...569A..38S} {569, A38}

\bibitem[\protect\citeauthoryear{Skilling}{Skilling}{2006}]{skilling2006}
Skilling J.,  2006, \mn@doi [Bayesian Anal.] {10.1214/06-BA127}, 1, 833

\bibitem[\protect\citeauthoryear{{Thompson} et~al.,}{{Thompson}
  et~al.}{2016}]{2016SPIE.9908E..6FT}
{Thompson} S.~J.,  et~al., 2016, in Ground-based and Airborne Instrumentation
  for Astronomy VI. p. 99086F

\bibitem[\protect\citeauthoryear{{Tuomi} et~al.,}{{Tuomi}
  et~al.}{2013}]{2013A&A...551A..79T}
{Tuomi} M.,  et~al., 2013, \mn@doi [\aap] {10.1051/0004-6361/201220509}, \href
  {http://adsabs.harvard.edu/abs/2013A%26A...551A..79T} {551, A79}

\bibitem[\protect\citeauthoryear{{Wang}, {Liu}, {Tian}, {Hu}  \&
  {Huang}}{{Wang} et~al.}{2017}]{2017arXiv171001405W}
{Wang} Y.,  {Liu} Y.,  {Tian} F.,  {Hu} Y.,   {Huang} Y.,  2017, preprint,
  \href {http://adsabs.harvard.edu/abs/2017arXiv171001405W} {} (\mn@eprint
  {arXiv} {1710.01405})

\bibitem[\protect\citeauthoryear{{Wheatley} et~al.,}{{Wheatley}
  et~al.}{2017}]{2017arXiv171011100W}
{Wheatley} P.~J.,  et~al., 2017, preprint, \href
  {http://adsabs.harvard.edu/abs/2017arXiv171011100W} {} (\mn@eprint {arXiv}
  {1710.11100})

\bibitem[\protect\citeauthoryear{{Winn} \& {Fabrycky}}{{Winn} \&
  {Fabrycky}}{2015}]{2015ARA&A..53..409W}
{Winn} J.~N.,  {Fabrycky} D.~C.,  2015, \mn@doi [\araa]
  {10.1146/annurev-astro-082214-122246}, \href
  {http://adsabs.harvard.edu/abs/2015ARA%26A..53..409W} {53, 409}

\makeatother
\end{thebibliography}



\appendix

\section{Tables of Results}
\label{app:results}

\begin{table*}
\centering
\caption{Model evidences and parameter estimates for System 1. \PolyChord{} is not constrained in the choice of model, only the parameters of only the favoured model are shown. For clarity the favoured model has been underlined. The data only contains planetary RVs and Gaussian noise. |$\Delta \ln R$| has the zero--point set to the favoured model. The correct model is indicated with a check mark. The results are discussed in section \ref{subsec:sys1_nosoap_results}.}
\label{tab:NO_SOAP_sys1}
\begin{tabular}{@{}lrrrrrlrrr@{}}
\cmidrule(r){1-6} \cmidrule(l){8-10}
 & \multicolumn{5}{c}{|$\Delta \ln R$|} &  &  &  &  \\
Schedule & $N_p = 0$ & $N_p = 1$ & $N_p = 2$ & $N_p = 3$ & $N_p = 4$ &  & \begin{tabular}[c]{@{}r@{}}RV Semi \\Amplitude / \\  \SI{}{\meter\per\second}\end{tabular} & \begin{tabular}[c]{@{}r@{}}Period / \\ days\end{tabular} & \begin{tabular}[c]{@{}r@{}}Phase / \\ radians\end{tabular} \\ \cmidrule(r){1-6} \cmidrule(l){8-10} 
True & \multicolumn{1}{l}{} & \multicolumn{1}{c}{\checkmark} & \multicolumn{1}{l}{} & \multicolumn{1}{l}{} & \multicolumn{1}{l}{} &  & 0.11 & 293 & 3.46 \\
REF\_5 & -4.48 & {\ul 0.00 $\pm$ 0.25} & -3.14 $\pm$ 0.20 & -6.16 $\pm$ 0.19 & -9.56 $\pm$ 0.18 &  & 0.13 $\pm$ 0.03 & 292.70 $\pm$ 82.17 & 3.23 $\pm$ 0.47 \\
REF\_10 & -9.00 & {\ul 0.00 $\pm$ 0.28} & -3.93 $\pm$ 0.23 & -6.70 $\pm$ 0.20 & -11.13 $\pm$ 0.21 &  & 0.11 $\pm$ 0.02 & 272.24 $\pm$ 70.07 & 3.08 $\pm$ 0.78 \\
THE\_5 & -18.68 & {\ul 0.00 $\pm$ 0.30} & -3.32 $\pm$ 0.24 & -6.27 $\pm$ 0.21 & -10.96 $\pm$ 0.23 &  & 0.11 $\pm$ 0.01 & 286.47 $\pm$ 7.45 & 2.88 $\pm$ 0.28 \\
THE\_10 & -37.24 & {\ul 0.00 $\pm$ 0.30} & -5.53 $\pm$ 0.24 & -8.76 $\pm$ 0.21 & -14.83 $\pm$ 0.23 &  & 0.10 $\pm$ 0.01 & 290.89 $\pm$ 1.25 & 3.13 $\pm$ 0.19 \\
Space\_5 & -51.24 & {\ul 0.00 $\pm$ 0.27} & -5.68 $\pm$ 0.26 & -9.30 $\pm$ 0.23 & -13.41 $\pm$ 0.22 &  & 0.11 $\pm$ 0.01 & 290.17 $\pm$ 2.16 & 3.28 $\pm$ 0.17 \\
Space\_10 & -109.84 & {\ul 0.00 $\pm$ 0.29} & -8.52 $\pm$ 0.29 & -11.42 $\pm$ 0.29 & -19.32 $\pm$ 0.26 &  & 0.11 $\pm$ 0.01 & 292.10 $\pm$ 0.08 & 3.41 $\pm$ 0.12 \\ \cmidrule(r){1-6} \cmidrule(l){8-10} 
\end{tabular}
\end{table*}
\begin{table*}
\centering
\caption{As per Table \ref{tab:NO_SOAP_sys1} for System 2, the results are discussed in section \ref{subsec:sys2_nosoap_results}.}
\label{tab:NO_SOAP_sys2}
\begin{tabular}{lrrrrrrrrr}
\cline{1-6} \cline{8-10}
 & \multicolumn{5}{c}{|$\Delta \ln R$|} & \multicolumn{1}{l}{} & \multicolumn{1}{l}{} & \multicolumn{1}{l}{} & \multicolumn{1}{l}{} \\
Schedule & $N_p = 0$ & $N_p = 1$ & $N_p = 2$ & $N_p = 3$ & $N_p = 4$ & \multicolumn{1}{l}{} & \begin{tabular}[c]{@{}r@{}}RV Semi\\ Amplitude / \\ $ms^-1$\end{tabular} & \begin{tabular}[c]{@{}r@{}}Period / \\ days\end{tabular} & \begin{tabular}[c]{@{}r@{}}Phase / \\ radians\end{tabular} \\ \cline{1-6} \cline{8-10} 
True &  &  &  & \multicolumn{1}{c}{\checkmark} &  & \multicolumn{1}{l}{} & 0.11 & 197 & 3.46 \\
 &  &  &  &  &  & \multicolumn{1}{l}{} & 0.11 & 293 & 4.45 \\
 &  &  &  &  &  & \multicolumn{1}{l}{} & 10.34 & 2953 & 1.83\smallskip\\
REF\_5 & -89,216.94 & -1.61 $\pm$ 0.37 & {\ul 0.00 $\pm$ 0.28} & -0.58 $\pm$ 0.28 & -1.11 $\pm$ 0.28 &  & 0.27 $\pm$ 0.50 & 864.46 $\pm$ 929.07 & 3.39 $\pm$ 1.43 \\
 &  &  &  &  &  &  & 10.33 $\pm$ 0.18 & 2947.92 $\pm$ 25.02 & 1.81 $\pm$ 0.04\smallskip\\
REF\_10 & -157,854.40 & -9.10 $\pm$ 0.38 & -2.93 $\pm$ 0.34 & -0.18 $\pm$ 0.38 & {\ul 0.00 $\pm$ 0.32} &  & 0.07 $\pm$ 0.02 & 159.37 $\pm$ 56.51 & 3.11 $\pm$ 0.92 \\
 &  &  &  &  &  &  & 0.09 $\pm$ 0.02 & 341.13 $\pm$ 378.18 & 3.66 $\pm$ 1.12 \\
 &  &  &  &  &  &  & 0.97 $\pm$ 2.92 & 1543.42 $\pm$ 565.01 & 1.33 $\pm$ 0.90 \\
 &  &  &  &  &  &  & 9.50 $\pm$ 2.82 & 2990.96 $\pm$ 138.74 & 2.14 $\pm$ 1.06\smallskip\\
THE\_5 & -266,916.05 & -41.56 $\pm$ 0.38 & -11.88 $\pm$ 0.38 & {\ul 0.00 $\pm$ 0.47} & -3.82 $\pm$ 0.46 &  & 0.12 $\pm$ 0.04 & 286.03 $\pm$ 117.44 & 2.89 $\pm$ 1.38 \\
 &  &  &  &  &  &  & 0.20 $\pm$ 0.38 & 767.69 $\pm$ 686.73 & 3.64 $\pm$ 1.68 \\
 &  &  &  &  &  &  & 10.30 $\pm$ 0.31 & 2963.98 $\pm$ 36.68 & 1.84 $\pm$ 0.13\smallskip\\
THE\_10 & -476,030.22 & -83.23 $\pm$ 0.40 & -31.29 $\pm$ 0.36 & {\ul 0.00 $\pm$ 0.73} & -1.49 $\pm$ 0.94 &  & 0.10 $\pm$ 0.01 & 197.45 $\pm$ 1.18 & 3.58 $\pm$ 0.20 \\
 &  &  &  &  &  &  & 0.12 $\pm$ 0.01 & 294.78 $\pm$ 1.96 & 4.77 $\pm$ 0.20 \\
 &  &  &  &  &  &  & 10.35 $\pm$ 0.01 & 2954.01 $\pm$ 1.44 & 1.83 $\pm$ 0.01\smallskip\\
Space\_5 & -572,825.73 & -102.19 $\pm$ 0.38 & -45.38 $\pm$ 0.39 & {\ul 0.00 $\pm$ 0.42} & -2.08 $\pm$ 0.52 &  & 0.10 $\pm$ 0.01 & 198.57 $\pm$ 1.14 & 3.62 $\pm$ 0.19 \\
 &  &  &  &  &  &  & 0.11 $\pm$ 0.01 & 292.82 $\pm$ 2.33 & 4.51 $\pm$ 0.17 \\
 &  &  &  &  &  &  & 10.35 $\pm$ 0.01 & 2957.53 $\pm$ 3.12 & 1.83 $\pm$ 0.01\smallskip\\
Space\_10 & -1,026,898.77 & -200.65 $\pm$ 0.40 & -89.99 $\pm$ 0.42 & {\ul 0.00 $\pm$ 0.48} & -1.49 $\pm$ 0.86 &  & 0.10 $\pm$ 0.01 & 196.66 $\pm$ 0.41 & 3.37 $\pm$ 0.14 \\
 &  &  &  &  &  &  & 0.11 $\pm$ 0.01 & 293.27 $\pm$ 0.81 & 4.58 $\pm$ 0.13 \\
 &  &  &  &  &  &  & 10.34 $\pm$ 0.01 & 2955.01 $\pm$ 0.84 & 1.83 $\pm$ 0.01 \\ \cline{1-6} \cline{8-10} 
\end{tabular}
\end{table*}
\begin{table*}
\centering
\caption{As per Table \ref{tab:NO_SOAP_sys1} for System 3, the results are discussed in section \ref{subsec:sys3_nosoap_results}.}
\label{tab:NO_SOAP_sys3}
\begin{tabular}{@{}lrrrrrlrrr@{}}
\cmidrule(r){1-6} \cmidrule(l){8-10}
 & \multicolumn{5}{c}{|$\Delta \ln R$|} &  & \multicolumn{1}{l}{} & \multicolumn{1}{l}{} & \multicolumn{1}{l}{} \\
Schedule & $N_p = 0$ & $N_p = 1$ & $N_p = 2$ & $N_p = 3$ & $N_p = 4$ &  & \begin{tabular}[c]{@{}r@{}}RV Semi \\ Amplitude / \\\SI{}{\meter\per\second}\end{tabular} & \begin{tabular}[c]{@{}r@{}}Period / \\ days\end{tabular} & \begin{tabular}[c]{@{}r@{}}Phase / \\ radians\end{tabular} \\ \cmidrule(r){1-6} \cmidrule(l){8-10} 
True &  &  &  & \multicolumn{1}{c}{\checkmark} &  &  & 0.16 & 101 & 3.46 \\
 &  &  &  &  &  &  & 0.13 & 197 & 4.45 \\
 &  &  &  &  &  &  & 0.11 & 293 & 1.83\smallskip \\
REF\_5 & -11.75 & -8.09 $\pm$ 0.27 & -1.57 $\pm$ 0.25 & -0.75 $\pm$ 0.33 & {\ul 0.00 $\pm$ 0.30} &  & 0.16 $\pm$ 0.04 & 99.08 $\pm$ 9.87 & 3.30 $\pm$ 0.46 \\
 & \multicolumn{1}{l}{} & \multicolumn{1}{l}{} & \multicolumn{1}{l}{} & \multicolumn{1}{l}{} & \multicolumn{1}{l}{} &  & 0.10 $\pm$ 0.03 & 167.30 $\pm$ 38.65 & 3.83 $\pm$ 1.67 \\
 & \multicolumn{1}{l}{} & \multicolumn{1}{l}{} & \multicolumn{1}{l}{} & \multicolumn{1}{l}{} & \multicolumn{1}{l}{} &  & 0.12 $\pm$ 0.03 & 248.01 $\pm$ 51.82 & 3.52 $\pm$ 1.76 \\
 & \multicolumn{1}{l}{} & \multicolumn{1}{l}{} & \multicolumn{1}{l}{} & \multicolumn{1}{l}{} & \multicolumn{1}{l}{} &  & 0.11 $\pm$ 0.03 & 508.95 $\pm$ 231.93 & 3.33 $\pm$ 1.53\smallskip \\
REF\_10 & -27.01 & -18.74 $\pm$ 0.27 & -6.07 $\pm$ 0.45 & -1.91 $\pm$ 0.35 & {\ul 0.00 $\pm$ 0.33} &  & 0.16 $\pm$ 0.02 & 74.80 $\pm$ 28.77 & 4.72 $\pm$ 1.14 \\
 & \multicolumn{1}{l}{} & \multicolumn{1}{l}{} & \multicolumn{1}{l}{} & \multicolumn{1}{l}{} & \multicolumn{1}{l}{} &  & 0.12 $\pm$ 0.02 & 196.61 $\pm$ 1.52 & 4.53 $\pm$ 0.54 \\
 & \multicolumn{1}{l}{} & \multicolumn{1}{l}{} & \multicolumn{1}{l}{} & \multicolumn{1}{l}{} & \multicolumn{1}{l}{} &  & 0.09 $\pm$ 0.02 & 272.01 $\pm$ 36.15 & 2.30 $\pm$ 1.15 \\
 & \multicolumn{1}{l}{} & \multicolumn{1}{l}{} & \multicolumn{1}{l}{} & \multicolumn{1}{l}{} & \multicolumn{1}{l}{} &  & 0.07 $\pm$ 0.02 & 526.09 $\pm$ 195.16 & 2.91 $\pm$ 1.48\smallskip \\
THE\_5 & -55.79 & -42.65 $\pm$ 0.29 & -17.69 $\pm$ 0.29 & {\ul 0.00 $\pm$ 0.36} & -1.21 $\pm$ 0.27 &  & 0.15 $\pm$ 0.01 & 100.83 $\pm$ 0.29 & 3.46 $\pm$ 0.19 \\
 &  &  &  &  &  &  & 0.12 $\pm$ 0.02 & 196.18 $\pm$ 1.48 & 4.40 $\pm$ 0.25 \\
 &  &  &  &  &  &  & 0.11 $\pm$ 0.02 & 294.97 $\pm$ 3.76 & 1.85 $\pm$ 0.27\smallskip \\
THE\_10 & -139.02 & -104.55 $\pm$ 0.30 & -44.00 $\pm$ 0.35 & {\ul 0.00 $\pm$ 0.36} & -9.70 $\pm$ 0.97 &  & 0.15 $\pm$ 0.01 & 101.14 $\pm$ 0.10 & 3.59 $\pm$ 0.14 \\
 &  &  &  &  &  &  & 0.12 $\pm$ 0.01 & 196.73 $\pm$ 0.51 & 4.47 $\pm$ 0.17 \\
 &  &  &  &  &  &  & 0.11 $\pm$ 0.01 & 292.32 $\pm$ 1.24 & 1.71 $\pm$ 0.20\smallskip \\
Space\_5 & -213.33 & -124.11 $\pm$ 0.31 & -52.09 $\pm$ 0.46 & {\ul 0.00 $\pm$ 0.30} & -2.14 $\pm$ 0.38 &  & 0.16 $\pm$ 0.01 & 100.66 $\pm$ 0.18 & 3.30 $\pm$ 0.12 \\
 & \multicolumn{1}{l}{} & \multicolumn{1}{l}{} & \multicolumn{1}{l}{} & \multicolumn{1}{l}{} & \multicolumn{1}{l}{} &  & 0.13 $\pm$ 0.01 & 195.86 $\pm$ 0.90 & 4.32 $\pm$ 0.15 \\
 & \multicolumn{1}{l}{} & \multicolumn{1}{l}{} & \multicolumn{1}{l}{} & \multicolumn{1}{l}{} & \multicolumn{1}{l}{} &  & 0.11 $\pm$ 0.01 & 292.81 $\pm$ 2.32 & 1.78 $\pm$ 0.17\smallskip \\
Space\_10 & -444.34 & -258.80 $\pm$ 0.33 & -118.82 $\pm$ 0.50 & {\ul 0.00 $\pm$ 0.30} & -18.99 $\pm$ 0.66 &  & 0.16 $\pm$ 0.01 & 100.98 $\pm$ 0.07 & 3.43 $\pm$ 0.09 \\
 &  &  &  &  &  &  & 0.12 $\pm$ 0.01 & 196.94 $\pm$ 0.33 & 4.46 $\pm$ 0.11 \\
 &  &  &  &  &  &  & 0.11 $\pm$ 0.01 & 292.46 $\pm$ 0.80 & 1.78 $\pm$ 0.12\smallskip \\ \cmidrule(r){1-6} \cmidrule(l){8-10} 
\end{tabular}
\end{table*}
\begin{table*}
\centering
\caption{As per Table \ref{tab:NO_SOAP_sys1} for System 4, the results are discussed in section \ref{subsec:sys4_nosoap_results}.}
\label{tab:NO_SOAP_sys4}
\begin{tabular}{lrrrrrlrrr}
\cline{1-6} \cline{8-10}
 & \multicolumn{5}{c}{|$\Delta \ln R$|} &  & \multicolumn{1}{l}{} & \multicolumn{1}{l}{} & \multicolumn{1}{l}{} \\
Schedule & $N_p = 0$ & $N_p = 1$ & $N_p = 2$ & $N_p = 3$ & $N_p = 4$ &  & \begin{tabular}[c]{@{}r@{}}RV Semi \\ Amplitude / \SI{}{\meter\per\second}\end{tabular} & \begin{tabular}[c]{@{}r@{}}Period / \\ days\end{tabular} & \begin{tabular}[c]{@{}r@{}}Phase / \\ radians\end{tabular} \\ \cline{1-6} \cline{8-10} 
True & \multicolumn{1}{c}{\checkmark} &  &  &  &  &  &  &  &  \\
REF\_5 & {\ul 0.00} & -27.43 $\pm$ 0.20 & -30.89 $\pm$ 0.19 & -33.64 $\pm$ 0.18 & -37.18 $\pm$ 0.17 &  & -- & -- & -- \\
REF\_10 & {\ul 0.00} & -41.00 $\pm$ 0.20 & -44.85 $\pm$ 0.19 & -47.83 $\pm$ 0.19 & -51.17 $\pm$ 0.19 &  & -- & -- & -- \\
THE\_5 & -3.95 & {\ul 0.00 $\pm$ 0.25} & -3.32 $\pm$ 0.24 & -8.90 $\pm$ 0.22 & -13.09 $\pm$ 0.20 &  & 0.06 $\pm$ 0.01 & 162.80 $\pm$ 107.97 & 2.21 $\pm$ 1.52 \\
THE\_10 & -3.27 & {\ul 0.00 $\pm$ 0.26} & -4.35 $\pm$ 0.22 & -10.45 $\pm$ 0.21 & -13.90 $\pm$ 0.31 &  & 0.05 $\pm$ 0.01 & 240.09 $\pm$ 155.15 & 3.13 $\pm$ 2.42 \\
Space\_5 & -35.18 & {\ul 0.00 $\pm$ 0.25} & -4.94 $\pm$ 0.23 & -10.08 $\pm$ 0.12 & -15.23 $\pm$ 0.20 &  & 0.05 $\pm$ 0.01 & 169.66 $\pm$ 50.58 & 3.20 $\pm$ 0.89 \\
Space\_10 & -9.89 & {\ul 0.00 $\pm$ 0.28} & -1.11 $\pm$ 0.22 & -8.59 $\pm$ 0.21 & -12.54 $\pm$ 0.23 &  & 0.05 $\pm$ 0.01 & 52.74 $\pm$ 56.10 & 3.65 $\pm$ 1.10 \\ \cline{1-6} \cline{8-10} 
\end{tabular}
\end{table*}

\begin{landscape}
\begin{table}
\centering
\caption{Parameter estimates for all schedules observing the first three Systems where the data includes planetary RVs, Gaussian noise and quasi--periodic stellar noise. \PolyChord{} has been constrained to the true number of planets. The results are discussed in section \ref{sec:constrained_results}.}
\label{table:soap_constrainted_full_results}
\begin{tabular}{lrrrlrrrlrrr}
\cline{1-4} \cline{6-8} \cline{10-12}
 & \multicolumn{3}{c}{System 1} &  & \multicolumn{3}{c}{System 2} &  & \multicolumn{3}{c}{System 3} \\
Schedule \smallskip & \begin{tabular}[c]{@{}r@{}}RV Semi Amplitude / \\ $ms^-1$\end{tabular} & \begin{tabular}[c]{@{}r@{}}Period / \\ days\end{tabular} & \begin{tabular}[c]{@{}r@{}}Phase / \\ radians\end{tabular} &  & \begin{tabular}[c]{@{}r@{}}RV Semi Amplitude / \\ $ms^-1$\end{tabular} & \begin{tabular}[c]{@{}r@{}}Period / \\ days\end{tabular} & \begin{tabular}[c]{@{}r@{}}Phase / \\ radians\end{tabular} &  & \begin{tabular}[c]{@{}r@{}}RV Semi Amplitude / \\ $ms^-1$\end{tabular} & \begin{tabular}[c]{@{}r@{}}Period / \\ days\end{tabular} & \begin{tabular}[c]{@{}r@{}}Phase / \\ radians\end{tabular} \\  \cline{1-4} \cline{6-8} \cline{10-12} 
True & 0.11 & 293 & 3.46 &  & 0.11 & 197 & 3.46 &  & 0.16 & 101 & 3.46 \\
 &  &  &  &  & 0.11 & 293 & 4.45 &  & 0.13 & 197 & 4.45 \\
 &  &  &  &  & 10.34 & 2953 & 1.82 &  & 0.11 & 293 & 1.82\smallskip \\
REF\_5 & 0.16 $\pm$ 0.03 & 210.13 $\pm$ 69.27 & 1.14 $\pm$ 0.42 &  & 0.11 $\pm$ 0.03 & 49.20 $\pm$ 28.09 & 2.03 $\pm$ 1.32 &  & 0.20 $\pm$ 0.03 & 76.25 $\pm$ 11.29 & 2.32 $\pm$ 2.77 \\
 &  &  &  &  & 0.19 $\pm$ 0.11 & 267.74 $\pm$ 295.75 & 5.88 $\pm$ 0.92 &  & 0.21 $\pm$ 0.05 & 203.33 $\pm$ 25.85 & 0.64 $\pm$ 1.33 \\
 &  &  &  &  & 10.38 $\pm$ 0.10 & 2956.10 $\pm$ 23.78 & 1.84 $\pm$ 0.01 &  & 0.12 $\pm$ 0.05 & 636.77 $\pm$ 470.34 & 2.46 $\pm$ 1.42\smallskip \\
REF\_10 & 0.12 $\pm$ 0.02 & 302.69 $\pm$ 65.13 & 3.01 $\pm$ 0.44 &  & 0.10 $\pm$ 0.02 & 27.50 $\pm$ 16.94 & 2.87 $\pm$ 0.97 &  & 0.16 $\pm$ 0.02 & 51.32 $\pm$ 24.55 & 1.20 $\pm$ 1.81 \\
 &  &  &  &  & 0.10 $\pm$ 0.41 & 198.67 $\pm$ 175.24 & 1.29 $\pm$ 0.85 &  & 0.16 $\pm$ 0.02 & 342.78 $\pm$ 133.52 & 2.92 $\pm$ 0.96 \\
 &  &  &  &  & 10.33 $\pm$ 0.41 & 2968.59 $\pm$ 30.95 & 1.85 $\pm$ 0.06 &  & 0.14 $\pm$ 0.02 & 1250.58 $\pm$ 384.32 & 3.95 $\pm$ 0.92\smallskip \\
THE\_5 & 0.13 $\pm$ 0.02 & 283.94 $\pm$ 3.01 & 2.44 $\pm$ 0.22 &  & 0.10 $\pm$ 0.02 & 20.88 $\pm$ 2.05 & 5.71 $\pm$ 0.42 &  & 0.15 $\pm$ 0.01 & 100.88 $\pm$ 0.35 & 3.25 $\pm$ 0.21 \\
 &  &  &  &  & 0.11 $\pm$ 0.02 & 206.65 $\pm$ 9.69 & 4.69 $\pm$ 0.30 &  & 0.10 $\pm$ 0.02 & 203.30 $\pm$ 12.47 & 5.26 $\pm$ 0.55 \\
 &  &  &  &  & 10.36 $\pm$ 0.02 & 2944.64 $\pm$ 4.66 & 1.83 $\pm$ 0.01 &  & 0.14 $\pm$ 0.02 & 294.94 $\pm$ 22.42 & 1.81 $\pm$ 0.30\smallskip \\
THE\_10 & 0.11 $\pm$ 0.01 & 290.76 $\pm$ 1.08 & 2.77 $\pm$ 0.17 &  & 0.07 $\pm$ 0.01 & 80.38 $\pm$ 33.64 & 4.70 $\pm$ 1.30 &  & 0.15 $\pm$ 0.01 & 100.88 $\pm$ 0.10 & 3.25 $\pm$ 0.14 \\
 &  &  &  &  & 0.08 $\pm$ 0.01 & 291.19 $\pm$ 18.56 & 4.45 $\pm$ 0.43 &  & 0.10 $\pm$ 0.01 & 197.58 $\pm$ 7.39 & 4.85 $\pm$ 0.30 \\
 &  &  &  &  & 10.35 $\pm$ 0.01 & 2959.08 $\pm$ 1.43 & 1.84 $\pm$ 0.01 &  & 0.14 $\pm$ 0.01 & 292.89 $\pm$ 10.46 & 1.70 $\pm$ 0.14\smallskip \\
Space\_5 & 0.12 $\pm$ 0.01 & 284.04 $\pm$ 2.12 & 2.67 $\pm$ 0.15 &  & 0.11 $\pm$ 0.01 & 202.20 $\pm$ 1.05 & 4.09 $\pm$ 0.19 &  & 0.16 $\pm$ 0.01 & 100.64 $\pm$ 0.20 & 3.23 $\pm$ 0.13 \\
 &  &  &  &  & 0.09 $\pm$ 0.01 & 288.47 $\pm$ 3.78 & 4.01 $\pm$ 0.31 &  & 0.11 $\pm$ 0.01 & 200.29 $\pm$ 1.08 & 4.89 $\pm$ 0.16 \\
 &  &  &  &  & 10.36 $\pm$ 0.01 & 2958.52 $\pm$ 3.08 & 1.83 $\pm$ 0.01 &  & 0.13 $\pm$ 0.01 & 291.80 $\pm$ 1.83 & 1.75 $\pm$ 0.13\smallskip \\
Space\_10 & 0.10 $\pm$ 0.01 & 291.73 $\pm$ 0.80 & 3.07 $\pm$ 0.13 &  & 0.09 $\pm$ 0.01 & 196.99 $\pm$ 0.48 & 3.31 $\pm$ 0.17 &  & 0.06 $\pm$ 0.01 & 23.05 $\pm$ 0.80 & 3.54 $\pm$ 0.23 \\
 &  &  &  &  & 0.08 $\pm$ 0.01 & 293.26 $\pm$ 1.27 & 4.49 $\pm$ 0.19 &  & 0.16 $\pm$ 0.01 & 100.86 $\pm$ 0.18 & 3.34 $\pm$ 0.09 \\
 &  &  &  &  & 10.34 $\pm$ 0.01 & 2955.60 $\pm$ 0.84 & 1.83 $\pm$ 0.01 &  & 0.14 $\pm$ 0.01 & 292.02 $\pm$ 0.77 & 1.67 $\pm$ 0.10 \\ \cline{1-4} \cline{6-8} \cline{10-12} 
\end{tabular}
\end{table}
\end{landscape}

\begin{table*}
\centering
\caption{Model evidences and parameter estimates for System 1, per observation schedule, with the data containing Stellar Noise, Gaussian noise and planetary RVs. \PolyChord{} has not been constrained to any model. The results are discussed in section \ref{subsec:sys1_soap_unforced}.}
\label{tab:sys1_unforced_results}
\begin{tabular}{lrrrrrlrrr}
\cline{1-6} \cline{8-10}
 & \multicolumn{5}{c}{|$\Delta \ln R$|} &  & \multicolumn{1}{l}{} & \multicolumn{1}{l}{} & \multicolumn{1}{l}{} \\
Schedule & $N_p = 0$ & $N_p = 1$ & $N_p = 2$ & $N_p = 3$ & $N_p = 4$ &  & \begin{tabular}[c]{@{}r@{}}RV Semi Amplitude / \\ \SI{}{\meter\per\second}\end{tabular} & \begin{tabular}[c]{@{}r@{}}Period / \\ days\end{tabular} & \begin{tabular}[c]{@{}r@{}}Phase / \\ radians\end{tabular} \\ \cline{1-6} \cline{8-10} 
True & \multicolumn{1}{l}{} & \multicolumn{1}{c}{\checkmark} & \multicolumn{1}{l}{} & \multicolumn{1}{l}{} & \multicolumn{1}{l}{} &  & 0.11 & 293 & 3.46\smallskip \\
REF\_5 & -19.14 & -8.75 $\pm$ 0.26 & -5.87 $\pm$ 0.24 & -3.05 $\pm$ 0.29 & {\ul 0.00 $\pm$ 0.23} &  & 0.12 $\pm$ 0.03 & 25.63 $\pm$ 16.65 & 4.55 $\pm$ 1.68 \\
 &  &  &  &  &  &  & 0.12 $\pm$ 0.04 & 46.30 $\pm$ 51.69 & 1.38 $\pm$ 0.65 \\
 &  &  &  &  &  &  & 0.13 $\pm$ 0.03 & 236.81 $\pm$ 120.35 & 1.64 $\pm$ 1.10 \\
 &  &  &  &  &  &  & 0.11 $\pm$ 0.03 & 3399.24 $\pm$ 500.80 & 3.16 $\pm$ 0.38\smallskip \\
REF\_10 & -32.59 & -19.39 $\pm$ 0.29 & -7.97 $\pm$ 0.25 & -3.66 $\pm$ 0.27 & {\ul 0.00 $\pm$ 0.26} &  & 0.08 $\pm$ 0.02 & 77.30 $\pm$ 60.32 & 3.11 $\pm$ 1.76 \\
 & \multicolumn{1}{l}{} & \multicolumn{1}{l}{} & \multicolumn{1}{l}{} & \multicolumn{1}{l}{} & \multicolumn{1}{l}{} &  & 0.10 $\pm$ 0.04 & 223.14 $\pm$ 67.84 & 1.64 $\pm$ 0.87 \\
 & \multicolumn{1}{l}{} & \multicolumn{1}{l}{} & \multicolumn{1}{l}{} & \multicolumn{1}{l}{} & \multicolumn{1}{l}{} &  & 0.14 $\pm$ 0.04 & 425.90 $\pm$ 181.64 & 3.04 $\pm$ 1.53 \\
 & \multicolumn{1}{l}{} & \multicolumn{1}{l}{} & \multicolumn{1}{l}{} & \multicolumn{1}{l}{} & \multicolumn{1}{l}{} &  & 0.11 $\pm$ 0.03 & 1344.35 $\pm$ 503.66 & 4.38 $\pm$ 1.84\smallskip \\
THE\_5 & -54.97 & -25.70 $\pm$ 0.28 & -15.80 $\pm$ 0.26 & -8.83 $\pm$ 0.36 & {\ul 0.00 $\pm$ 0.44} &  & 0.08 $\pm$ 0.01 & 17.98 $\pm$ 2.40 & 2.61 $\pm$ 1.69 \\
 & \multicolumn{1}{l}{} & \multicolumn{1}{l}{} & \multicolumn{1}{l}{} & \multicolumn{1}{l}{} & \multicolumn{1}{l}{} &  & 0.09 $\pm$ 0.02 & 24.80 $\pm$ 11.11 & 1.37 $\pm$ 0.50 \\
 & \multicolumn{1}{l}{} & \multicolumn{1}{l}{} & \multicolumn{1}{l}{} & \multicolumn{1}{l}{} & \multicolumn{1}{l}{} &  & 0.13 $\pm$ 0.02 & 285.30 $\pm$ 6.94 & 2.61 $\pm$ 0.27 \\
 & \multicolumn{1}{l}{} & \multicolumn{1}{l}{} & \multicolumn{1}{l}{} & \multicolumn{1}{l}{} & \multicolumn{1}{l}{} &  & 0.09 $\pm$ 0.02 & 395.91 $\pm$ 14.34 & 3.12 $\pm$ 0.36\smallskip \\
THE\_10 & -63.48 & -18.48 $\pm$ 0.30 & -7.19 $\pm$ 0.43 & {\ul 0.00 $\pm$ 0.48} & -0.48 $\pm$ 0.35 &  & 0.07 $\pm$ 0.01 & 60.64 $\pm$ 19.03 & 1.44 $\pm$ 2.22 \\
 & \multicolumn{1}{l}{} & \multicolumn{1}{l}{} & \multicolumn{1}{l}{} & \multicolumn{1}{l}{} & \multicolumn{1}{l}{} &  & 0.08 $\pm$ 0.02 & 217.82 $\pm$ 39.22 & 0.98 $\pm$ 0.93 \\
 & \multicolumn{1}{l}{} & \multicolumn{1}{l}{} & \multicolumn{1}{l}{} & \multicolumn{1}{l}{} & \multicolumn{1}{l}{} &  & 0.10 $\pm$ 0.02 & 453.28 $\pm$ 343.07 & 3.27 $\pm$ 1.09\smallskip \\
Space\_5 & -93.17 & -34.77 $\pm$ 0.28 & -14.56 $\pm$ 0.30 & -6.27 $\pm$ 0.48 & {\ul 0.00 $\pm$ 0.41} &  & 0.08 $\pm$ 0.01 & 21.98 $\pm$ 0.02 & 1.30 $\pm$ 0.29 \\
 &  &  &  &  &  &  & 0.06 $\pm$ 0.01 & 27.47 $\pm$ 9.65 & 1.91 $\pm$ 0.46 \\
 &  &  &  &  &  &  & 0.06 $\pm$ 0.01 & 37.70 $\pm$ 18.49 & 2.91 $\pm$ 0.29 \\
 &  &  &  &  &  &  & 0.12 $\pm$ 0.01 & 283.85 $\pm$ 8.96 & 2.65 $\pm$ 0.16\smallskip \\
Space\_10 & -132.44 & -37.66 $\pm$ 0.30 & -22.15 $\pm$ 0.32 & -9.98 $\pm$ 0.34 & {\ul 0.00 $\pm$ 0.53} &  & 0.06 $\pm$ 0.01 & 22.13 $\pm$ 0.33 & 2.84 $\pm$ 0.41 \\
 & \multicolumn{1}{l}{} & \multicolumn{1}{l}{} & \multicolumn{1}{l}{} & \multicolumn{1}{l}{} & \multicolumn{1}{l}{} &  & 0.05 $\pm$ 0.01 & 26.00 $\pm$ 0.15 & 6.06 $\pm$ 0.32 \\
 & \multicolumn{1}{l}{} & \multicolumn{1}{l}{} & \multicolumn{1}{l}{} & \multicolumn{1}{l}{} & \multicolumn{1}{l}{} &  & 0.06 $\pm$ 0.01 & 109.24 $\pm$ 0.25 & 1.51 $\pm$ 0.28 \\
 & \multicolumn{1}{l}{} & \multicolumn{1}{l}{} & \multicolumn{1}{l}{} & \multicolumn{1}{l}{} & \multicolumn{1}{l}{} &  & 0.10 $\pm$ 0.01 & 291.90 $\pm$ 0.84 & 3.08 $\pm$ 0.13 \\ \cline{1-6} \cline{8-10} 
\end{tabular}
\end{table*}


\begin{table*}
\centering
\caption{As in Table\ref{tab:sys1_unforced_results} but for System 2. The results are discussed in section \ref{subsec:sys2_soap_unforced}}
\label{tab:sys2_unforced_results}
\begin{tabular}{lrrrrrlrrr}
\cline{1-6} \cline{8-10}
 & \multicolumn{5}{c}{|$\Delta \ln R$|} &  & \multicolumn{1}{l}{} & \multicolumn{1}{l}{} & \multicolumn{1}{l}{} \\
Schedule & $N_p = 0$ & $N_p = 1$ & $N_p = 2$ & $N_p = 3$ & $N_p = 4$ &  & \begin{tabular}[c]{@{}r@{}}RV Semi \\Amplitude / \\ \SI{}{\meter\per\second}\end{tabular} & \begin{tabular}[c]{@{}r@{}}Period / \\ days\end{tabular} & \begin{tabular}[c]{@{}r@{}}Phase / \\ radians\end{tabular} \\ \cline{1-6} \cline{8-10} 
True &  &  &  &  \multicolumn{1}{c}{\checkmark} &  &  & 0.11 & 197 & 3.46 \\
 &  &  &  &  &  &  & 0.11 & 293 & 4.45 \\
 &  &  &  &  &  &  & 10.34 & 2953 & 1.83\smallskip\\
REF\_5 & -89,071.19 & -17.39 $\pm$ 0.36 & -4.92 $\pm$ 0.33 & -0.79 $\pm$ 0.37 & {\ul 0.00 $\pm$ 0.33} &  & 0.13 $\pm$ 0.04 & 44.32 $\pm$ 26.74 & 2.11 $\pm$ 1.36 \\
 &  &  &  &  &  &  & 0.15 $\pm$ 0.04 & 168.51 $\pm$ 117.21 & 4.62 $\pm$ 1.94 \\
 &  &  &  &  &  &  & 0.31 $\pm$ 1.28 & 1541.14 $\pm$ 402.76 & 1.20 $\pm$ 0.87 \\
 &  &  &  &  &  &  & 10.17 $\pm$ 1.27 & 2995.67 $\pm$ 76.40 & 1.86 $\pm$ 0.08\smallskip\\
REF\_10 & -157,356.78 & -9.20 $\pm$ 0.38 & -4.90 $\pm$ 0.42 & {\ul 0.00 $\pm$ 0.34} & -0.92 $\pm$ 0.44 &  & 0.10 $\pm$ 0.02 & 27.50 $\pm$ 16.94 & 2.87 $\pm$ 0.97 \\
 &  &  &  &  &  &  & 0.10 $\pm$ 0.41 & 198.67 $\pm$ 175.24 & 1.29 $\pm$ 0.85 \\
 &  &  &  &  &  &  & 10.33 $\pm$ 0.41 & 2968.59 $\pm$ 30.95 & 1.85 $\pm$ 0.06\smallskip\\
THE\_5 & -267,224.20 & -38.88 $\pm$ 0.37 & -20.60 $\pm$ 0.37 & -7.85 $\pm$ 0.39 & {\ul 0.00 $\pm$ 0.88} &  & 0.10 $\pm$ 0.01 & 22.05 $\pm$ 0.50 & 1.60 $\pm$ 0.38 \\
 &  &  &  &  &  &  & 0.10 $\pm$ 0.02 & 36.28 $\pm$ 46.11 & 0.96 $\pm$ 0.57 \\
 &  &  &  &  &  &  & 0.10 $\pm$ 0.09 & 321.02 $\pm$ 42.21 & 5.17 $\pm$ 1.06 \\
 &  &  &  &  &  &  & 10.33 $\pm$ 0.09 & 2953.60 $\pm$ 9.79 & 1.83 $\pm$ 0.03\smallskip\\
THE\_10 & -475,737.51 & -42.91 $\pm$ 0.39 & -24.27 $\pm$ 0.33 & -10.93 $\pm$ 0.38 & {\ul 0.00 $\pm$ 0.56} &  & 0.07 $\pm$ 0.01 & 55.14 $\pm$ 31.41 & 2.87 $\pm$ 2.22 \\
 &  &  &  &  &  &  & 0.07 $\pm$ 0.01 & 173.14 $\pm$ 1.57 & 2.35 $\pm$ 0.48 \\
 &  &  &  &  &  &  & 0.08 $\pm$ 0.01 & 295.56 $\pm$ 4.42 & 4.68 $\pm$ 0.53 \\
 &  &  &  &  &  &  & 10.35 $\pm$ 0.01 & 2958.42 $\pm$ 1.97 & 1.84 $\pm$ 0.01\smallskip\\
Space\_5 & -572,795.30 & -101.15 $\pm$ 0.39 & -57.53 $\pm$ 0.38 & -21.51 $\pm$ 037 & {\ul 0.00 $\pm$ 0.38} &  & 0.08 $\pm$ 0.01 & 21.97 $\pm$ 0.92 & 1.03 $\pm$ 0.35 \\
 &  &  &  &  &  &  & 0.11 $\pm$ 0.01 & 202.20 $\pm$ 1.09 & 4.09 $\pm$ 0.19 \\
 &  &  &  &  &  &  & 0.09 $\pm$ 0.01 & 288.41 $\pm$ 3.82 & 4.01 $\pm$ 0.31 \\
 &  &  &  &  &  &  & 10.36 $\pm$ 0.01 & 2958.52 $\pm$ 3.03 & 1.83 $\pm$ 0.01\smallskip\\
Space\_10 & -1,027,062.81 & -130.12 $\pm$ 0.40 & -65.05 $\pm$ 0.36 & -13.89 $\pm$ 0.54 & {\ul 0.00 $\pm$ 1.01} &  & 0.05 $\pm$ 0.01 & 49.35 $\pm$ 54.71 & 2.86 $\pm$ 0.37 \\
 &  &  &  &  &  &  & 0.09 $\pm$ 0.01 & 196.95 $\pm$ 0.94 & 3.29 $\pm$ 0.18 \\
 & \multicolumn{1}{l}{} & \multicolumn{1}{l}{} & \multicolumn{1}{l}{} & \multicolumn{1}{l}{} & \multicolumn{1}{l}{} &  & 0.08 $\pm$ 0.01 & 293.23 $\pm$ 1.53 & 4.48 $\pm$ 0.18 \\
 & \multicolumn{1}{l}{} & \multicolumn{1}{l}{} & \multicolumn{1}{l}{} & \multicolumn{1}{l}{} & \multicolumn{1}{l}{} &  & 10.34 $\pm$ 0.01 & 2955.55 $\pm$ 0.82 & 1.83 $\pm$ 0.11 \\ \cline{1-6} \cline{8-10} 
\end{tabular}
\end{table*}


\begin{table*}
\centering
\caption{As in Table\ref{tab:sys1_unforced_results} but for System 3. The results are discussed in section \ref{subsec:sys3_soap_unforced}}
\label{tab:sys3_unforced_results}
\begin{tabular}{lrrrrrlrrr}
\cline{1-6} \cline{8-10}
 & \multicolumn{5}{c}{|$\Delta \ln R$|} &  & \multicolumn{1}{l}{} & \multicolumn{1}{l}{} & \multicolumn{1}{l}{} \\
Schedule & $N_p = 0$ & $N_p = 1$ & $N_p = 2$ & $N_p = 3$ & $N_p = 4$ &  & \begin{tabular}[c]{@{}r@{}}RV Semi \\ Amplitude / \\ \SI{}{\meter\per\second}\end{tabular} & \begin{tabular}[c]{@{}r@{}}Period / \\ days\end{tabular} & \begin{tabular}[c]{@{}r@{}}Phase / \\ radians\end{tabular} \\ \cline{1-6} \cline{8-10} 
True &  &  &  & \multicolumn{1}{c}{\checkmark} &  &  & 0.16 & 101 & 3.46 \\
 &  &  &  &  &  &  & 0.13 & 197 & 4.45 \\
 &  &  &  &  &  &  & 0.11 & 293 & 1.83\smallskip\\
REF\_5 & -35.96 & -25.37 $\pm$ 0.27 & -5.99 $\pm$ 0.40 & -3.24 $\pm$ 0.33 & {\ul 0.00 $\pm$ 0.45} &  & 0.17 $\pm$ 0.07 & 25.50 $\pm$ 8.39 & 2.31 $\pm$ 1.71 \\
 &  &  &  &  &  &  & 0.20 $\pm$ 0.03 & 84.74 $\pm$ 18.76 & 2.80 $\pm$ 2.85 \\
 &  &  &  &  &  &  & 0.16 $\pm$ 0.06 & 300.15 $\pm$ 101.78 & 4.35 $\pm$ 1.82 \\
 &  &  &  &  &  &  & 0.13 $\pm$ 0.04 & 903.27 $\pm$ 391.54 & 3.93 $\pm$ 0.58\smallskip\\
REF\_10 & -40.94 & -64.01 $\pm$ 0.29 & -33.49 $\pm$ 0.35 & -16.00 $\pm$ 0.42 & {\ul 0.00 $\pm$ 0.37} &  & 0.16 $\pm$ 0.02 & 36.70 $\pm$ 16.28 & 3.50 $\pm$ 2.85 \\
 &  &  &  &  &  &  & 0.15 $\pm$ 0.03 & 154.11 $\pm$ 61.49 & 4.34 $\pm$ 1.52 \\
 &  &  &  &  &  &  & 0.16 $\pm$ 0.05 & 338.09 $\pm$ 112.05 & 2.28 $\pm$ 1.17 \\
 &  &  &  &  &  & \multicolumn{1}{r}{} & 0.16 $\pm$ 0.03 & 1342.12 $\pm$ 207.09 & 3.87 $\pm$ 0.74\smallskip\\
THE\_5 & -81.69 & -72.10 $\pm$ 0.27 & -33.75 $\pm$ 0.29 & -21.61 $\pm$ 0.31 & {\ul 0.00 $\pm$ 0.28} &  & 0.11 $\pm$ 0.01 & 21.90 $\pm$ 0.29 & 1.54 $\pm$ 0.74 \\
 &  &  &  &  &  &  & 0.15 $\pm$ 0.01 & 100.87 $\pm$ 0.34 & 3.25 $\pm$ 0.21 \\
 &  &  &  &  &  &  & 0.10 $\pm$ 0.02 & 202.06 $\pm$ 2.23 & 5.32 $\pm$ 0.28 \\
 &  &  &  &  &  &  & 0.14 $\pm$ 0.01 & 292.43 $\pm$ 2.88 & 1.78 $\pm$ 0.21\smallskip\\
THE\_10 & -163.46 & -140.08 $\pm$ 0.30 & -57.63 $\pm$ 0.73 & -19.08 $\pm$ 0.43 & {\ul 0.00 $\pm$ 0.45} &  & 0.08 $\pm$ 0.01 & 26.03 $\pm$ 0.28 & 3.87 $\pm$ 2.81 \\
 &  &  &  &  &  &  & 0.15 $\pm$ 0.01 & 100.87 $\pm$ 0.10 & 3.24 $\pm$ 0.13 \\
 &  &  &  &  &  &  & 0.10 $\pm$ 0.01 & 196.96 $\pm$ 0.55 & 4.84 $\pm$ 0.19 \\
 &  &  &  &  &  &  & 0.14 $\pm$ 0.01 & 291.99 $\pm$ 0.93 & 1.68 $\pm$ 0.14\smallskip\\
Space\_5 & -246.16 & -150.70 $\pm$ 0.29 & -67.53 $\pm$ 0.32 & -15.77 $\pm$ 0.29 & {\ul 0.00 $\pm$ 0.42} &  & 0.11 $\pm$ 0.05 & 54.62 $\pm$ 38.23 & 2.61 $\pm$ 1.06 \\
 &  &  &  &  &  &  & 0.14 $\pm$ 0.03 & 140.49 $\pm$ 48.03 & 3.80 $\pm$ 0.71 \\
 &  &  &  &  &  &  & 0.10 $\pm$ 0.02 & 221.09 $\pm$ 25.28 & 3.38 $\pm$ 1.95 \\
 &  &  &  &  &  &  & 0.14 $\pm$ 0.02 & 289.50 $\pm$ 3.67 & 1.59 $\pm$ 0.27\smallskip\\
Space\_10 & -478.19 & -308.00 $\pm$ 0.33 & -113.51 $\pm$ 0.30 & -99.73 $\pm$ 0.47 & {\ul 0.00 $\pm$ 0.50} &  & 0.06 $\pm$ 0.01 & 23.04 $\pm$ 0.47 & 3.54 $\pm$ 0.23 \\
 &  &  &  &  &  &  & 0.16 $\pm$ 0.06 & 100.84 $\pm$ 0.06 & 3.34 $\pm$ 0.08 \\
 & \multicolumn{1}{l}{} & \multicolumn{1}{l}{} & \multicolumn{1}{l}{} & \multicolumn{1}{l}{} & \multicolumn{1}{l}{} &  & 0.11 $\pm$ 0.01 & 197.35 $\pm$ 0.35 & 4.62 $\pm$ 0.12 \\
 & \multicolumn{1}{l}{} & \multicolumn{1}{l}{} & \multicolumn{1}{l}{} & \multicolumn{1}{l}{} & \multicolumn{1}{l}{} &  & 0.14 $\pm$ 0.01 & 292.60 $\pm$ 0.60 & 1.74 $\pm$ 0.09 \\ \cline{1-6} \cline{8-10} 
\end{tabular}
\end{table*}

\begin{table*}
\centering
\caption{As in Table\ref{tab:sys1_unforced_results} but for System 4. The results are discussed in section \ref{subsec:sys4_soap_unforced}}
\label{tab:sys4_unforced_results}
\begin{tabular}{lrrrrrlrrr}
\cline{1-6} \cline{8-10}
 & \multicolumn{5}{c}{|$\Delta \ln R$|} &  & \multicolumn{1}{l}{} & \multicolumn{1}{l}{} & \multicolumn{1}{l}{} \\
Schedule & $N_p = 0$ & $N_p = 1$ & $N_p = 2$ & $N_p = 3$ & $N_p = 4$ &  & \begin{tabular}[c]{@{}r@{}}RV Semi \\ Amplitude / \\ \SI{}{\meter\per\second}\end{tabular} & \begin{tabular}[c]{@{}r@{}}Period / \\ days\end{tabular} & \begin{tabular}[c]{@{}r@{}}Phase / \\ radians\end{tabular} \\ \cline{1-6} \cline{8-10} 
True & \multicolumn{1}{l}{} & \multicolumn{1}{l}{} & \multicolumn{1}{l}{} & \multicolumn{1}{l}{} & \multicolumn{1}{l}{} &  & \multicolumn{1}{c}{N/A} & \multicolumn{1}{c}{N/A} & \multicolumn{1}{c}{N/A}\smallskip\\
REF\_5 & {\ul 0.00} & -3.14 $\pm$ 0.32 & -3.52 $\pm$ 0.22 & -5.01 $\pm$ 0.21 & -6.79 $\pm$ 0.20 &  &  &  &\smallskip\\
REF\_10 & {\ul 0.00} & -33.33 $\pm$ 0.23 & -28.32 $\pm$ 0.23 & -25.11 $\pm$ 0.24 & -19.73 $\pm$ 0.25 &  &  &  &\smallskip\\
THE\_5 & -40.08 & -10.13 $\pm$ 0.24 & -7.04 $\pm$ 0.24 & {\ul 0.00 $\pm$ 0.31} & -5.79 $\pm$ 0.27 &  & 0.07 $\pm$ 0.01 & 16.57 $\pm$ 1.14 & 0.85 $\pm$ 0.96 \\
 &  &  &  &  &  &  & 0.08 $\pm$ 0.01 & 23.98 $\pm$ 15.87 & 4.88 $\pm$ 0.58 \\
 &  &  &  &  &  &  & 0.08 $\pm$ 0.01 & 379.34 $\pm$ 39.94 & 2.10 $\pm$ 0.68\smallskip\\
THE\_10 & -19.43 & -9.90 $\pm$ 0.30 & -10.61 $\pm$ 0.28 & -2.89 $\pm$ 0.34 & {\ul 0.00 $\pm$ 0.43} &  & 0.07 $\pm$ 0.01 & 25.78 $\pm$ 1.05 & 2.77 $\pm$ 2.78 \\
 &  &  &  &  &  &  & 0.07 $\pm$ 0.01 & 30.74 $\pm$ 4.30 & 2.98 $\pm$ 2.67 \\
 &  &  &  &  &  &  & 0.07 $\pm$ 0.02 & 544.98 $\pm$ 412.99 & 2.99 $\pm$ 1.68 \\
 &  &  &  &  &  &  & 0.06 $\pm$ 0.02 & 1007.50 $\pm$ 233.98 & 3.37 $\pm$ 0.96\smallskip\\
Space\_5 & -110.53 & -19.56 $\pm$ 0.26 & -10.43 $\pm$ 0.35 & -3.19 $\pm$ 0.39 & {\ul 0.00 $\pm$ 0.38} &  & 0.06 $\pm$ 0.01 & 137.24 $\pm$ 14.58 & 4.94 $\pm$ 0.55 \\
 &  &  &  &  &  &  & 0.06 $\pm$ 0.01 & 180.91 $\pm$ 6.80 & 3.83 $\pm$ 0.41 \\
 &  &  &  &  &  &  & 0.06 $\pm$ 0.01 & 249.32 $\pm$ 8.62 & 0.80 $\pm$ 0.74 \\
 &  &  &  &  &  &  & 0.06 $\pm$ 0.01 & 384.23 $\pm$ 32.34 & 2.41 $\pm$ 0.67\smallskip\\
Space\_10 & -54.50 & -16.42 $\pm$ 0.32 & -10.74 $\pm$ 0.30 & {\ul 0.00 $\pm$ 0.30} & -7.48 $\pm$ .72 &  & 0.06 $\pm$ 0.01 & -22.11 $\pm$ 0.01 & 2.82 $\pm$ 0.27 \\
 &  &  &  &  &  &  & 0.05 $\pm$ 0.01 & 26.49 $\pm$ 0.28 & 1.29 $\pm$ 0.23 \\
 & \multicolumn{1}{l}{} & \multicolumn{1}{l}{} & \multicolumn{1}{l}{} & \multicolumn{1}{l}{} & \multicolumn{1}{l}{} &  & 0.05 $\pm$ 0.01 & 871.00 $\pm$ 15.61 & 1.93 $\pm$ 0.30 \\ \cline{1-6} \cline{8-10} 
\end{tabular}
\end{table*}


\bsp	
\label{lastpage}
\end{document}